%% file: network-test.tex
\documentclass{sig-alternate-10pt}
\usepackage[font=bf]{caption}

\newcommand{\subparagraph}{}

\usepackage[compact]{titlesec}

\usepackage{multirow}
\usepackage[sort,nocompress]{cite}
\usepackage[usenames,dvipsnames]{color}
\usepackage{amsfonts}
\usepackage{times}
\usepackage{xspace} \usepackage{epsfig} 
\usepackage{amsmath}
\usepackage{amssymb}
\usepackage[hyphens]{url}
\usepackage{listings}
\usepackage{fancyvrb}
\usepackage{graphicx}
\usepackage{subfig}
\usepackage[english]{babel}
\usepackage{courier}

\usepackage{listings}
\VerbatimFootnotes

\newcommand{\tightcaption}[1]{\vspace{-0.2cm}\caption{#1}\vspace{-0.4cm}}

\newcommand{\comment}[1]{}
\newcommand{\mycomment}[1]{}
\newcounter{note}[section]

\newcommand{\Section}{\S}

\usepackage{pifont}

\input{macros}

\input{macros_formulation}

\newcommand{\mypara}[1]{\smallskip\noindent{\bf {#1}:}~}
\newcommand{\myparatight}[1]{\smallskip\noindent{\bf {#1}:}~}
\newcommand{\myparaq}[1]{\smallskip\noindent{\bf {#1}?}~}

\newcommand{\vyas}[1]{}
\newcommand{\seyed}[1]{}

\newcounter{packednmbr}

\newenvironment{packedenumerate}{\begin{list}{\thepackednmbr.}{\usecounter{packednmbr}\setlength{\itemsep}{0.5pt}\addtolength{\labelwidth}{-4pt}\setlength{\leftmargin}{\labelwidth}\setlength{\listparindent}{\parindent}\setlength{\parsep}{1pt}\setlength{\topsep}{0pt}}}{\end{list}}

\newenvironment{packeditemize}{\begin{list}{$\bullet$}{\setlength{\itemsep}{0.5pt}\addtolength{\labelwidth}{-4pt}\setlength{\leftmargin}{\labelwidth}\setlength{\listparindent}{\parindent}\setlength{\parsep}{1pt}\setlength{\topsep}{0pt}}}{\end{list}}

\begin{document}

\title{Scalable Testing  of Context-Dependent Policies \\ over  Stateful Data Planes with \Name \vspace{-1pt}}

\author{Seyed K. Fayaz, Yoshiaki Tobioka, Sagar Chaki, Vyas Sekar \\ Carnegie Mellon University \vspace{-1pt}} 


\maketitle

\input{abstract} 
\input{intro} 
\input{motivation_vyas} 
\input{formulation}

\input{overview_vyas}  
\input{traffic_unit}  
\input{dpmodels}  
\input{testgenerate}  
\input{validation} 
\input{implementation} 
\input{evaluation_new} 
\input{relwork}  
\input{conclusions}

\newcommand\blfootnote[1]{%
  \begingroup
  \renewcommand\thefootnote{}\footnote{#1}%
  \addtocounter{footnote}{-1}%
  \endgroup
}

\renewcommand{\baselinestretch}{0.05}
\blfootnote{\fontsize{4}{11}\selectfont Copyright 2014 Carnegie Mellon
  University. This material is based upon
  work funded and supported by the Department of Defense under
  Contract No. FA8721-05-C-0003 with Carnegie Mellon University for
  the operation of the Software Engineering Institute, a federally
  funded research and development center. NO WARRANTY. THIS CARNEGIE
  MELLON UNIVERSITY AND SOFTWARE ENGINEERING INSTITUTE MATERIAL IS
  FURNISHED ON AN “AS-IS” BASIS. CARNEGIE MELLON UNIVERSITY MAKES NO
  WARRANTIES OF ANY KIND, EITHER EXPRESSED OR IMPLIED, AS TO ANY
  MATTER INCLUDING, BUT NOT LIMITED TO, WARRANTY OF FITNESS FOR
  PURPOSE OR MERCHANTABILITY, EXCLUSIVITY, OR RESULTS OBTAINED FROM
  USE OF THE MATERIAL. CARNEGIE MELLON UNIVERSITY DOES NOT MAKE ANY
  WARRANTY OF ANY KIND WITH RESPECT TO FREEDOM FROM PATENT, TRADEMARK,
  OR COPYRIGHT INFRINGEMENT. This material has been approved for
  public release and unlimited distribution. DM-0001673.}

\newpage
{
\footnotesize 
\bibliographystyle{abbrv}
\bibliography{bibs/seyed,bibs/mboxsdn,bibs/aplomb,bibs/mbox,bibs/service_chaining}
}


\end{document}

%% file: macros.tex
\newcommand{\Name}{{Armstrong}\xspace}

\newcommand{\klee}{\textsc{klee}\xspace}

\newcommand{\PADU}{ADU\xspace}
\newcommand{\PADUSeq}{ADUSeq\xspace}
\newcommand{\PADUs}{ADUs\xspace}
\newcommand{\testscript}{script\xspace}
\newcommand{\testscriptindex}{\ensuremath{\mathit{l}}\xspace}
\newcommand{\fsmensemble}{FSM ensemble\xspace}
\newcommand{\fsmensembles}{FSM ensembles\xspace}

\newcommand{\allPortsMon}{MonitorAll\xspace}
\newcommand{\dpfPortsMon}{MDP\xspace}

\newcommand{\cbmc}{\textsc{cmbc}\xspace}

\newcommand{\DPF}{NF\xspace}
\newcommand{\DPFs}{NFs\xspace}

\newcommand{\DPGs}{DPGs\xspace}

\newcommand{\intent}{policy\xspace}
\newcommand{\intents}{policies\xspace}

\newcommand{\Transfer}{\ensuremath{\mathit{T}}}
\newcommand{\header}{\ensuremath{\mathit{h}}}
\newcommand{\iport}{\ensuremath{\mathit{p}}}
\newcommand{\state}{\ensuremath{\mathit{s}}}
\newcommand{\context}{\ensuremath{\mathit{c}}}

\newcommand{\SE}{SE\xspace}

%% file: macros_formulation.tex
\newcommand{\set}[1]{\{{#1}\}}
\newcommand{\seq}[1]{\langle{#1}\rangle}
\newcommand{\such}{\ensuremath{\mathit{:\ }}}

\newcommand{\packet}{\mathcal{\pi}}
\newcommand{\inpacket}{\mathcal{\pi}^{\mathit{in}}}

\newcommand{\packets}{\mathcal{P}}
\newcommand{\ActionSet}{\ensuremath{\Sigma}}
\newcommand{\ActionVar}{\ensuremath{\alpha}}

\newcommand{\port}{port}
\newcommand{\edges}{edges\xspace}
\newcommand{\edge}{edge\xspace}

\newcommand{\action}{effect\xspace}
\newcommand{\actionTags}{c-Tags\xspace}
\newcommand{\actionTag}{c-Tag\xspace}
\newcommand{\actions}{effects\xspace}

\newcommand{\DPFState}{\ensuremath{\mathit{S}}}
\newcommand{\NumDPFs}{\ensuremath{\mathit{N}}}
\newcommand{\DPFStateVar}{\ensuremath{\mathit{s}}}
\newcommand{\DPFInit}{\ensuremath{\mathit{I}}}
\newcommand{\Ports}{\ensuremath{\mathit{E}}}
\newcommand{\PortVar}{\ensuremath{\mathit{e}}}
\newcommand{\Transition}{\ensuremath{\delta}}

\newcommand{\nodes}{\ensuremath{\mathit{N}}}
\newcommand{\PathLength}{\ensuremath{n}}
\newcommand{\NumPkts}{\ensuremath{m}}
\newcommand{\PktIndex}{\ensuremath{k}}
\newcommand{\topo}{\tau}
\newcommand{\netw}{\ensuremath{\mathit{net}}}

\newcommand{\DPFIndex}{\ensuremath{\mathit{i}}}
\newcommand{\sstep}[3]{{#1}\stackrel{#2}{\longrightarrow}{#3}}

\newcommand{\NetState}{\ensuremath{\sigma}}
\newcommand{\aseq}{\vec{\alpha}}
\newcommand{\etoesem}[1]{\ensuremath{\mathit{E2ESem}({#1}})}

\newcommand{\Policy}{policy\xspace}
\newcommand{\Trace}{\ensuremath{\Pi}}
\newcommand{\TraceSem}{\ensuremath{\mathit{TraceSem}}}
\newcommand{\TracePredicate}{\ensuremath{\mathit{TraceSpec}}}

\newcommand{\PreCondition}{\ensuremath{\ActionVar^\mathit{Before}}}
\newcommand{\PostCondition}{\ensuremath{\vec{\ActionVar}^\mathit{After}}}

\newcommand{\Action}{\ensuremath{\mathit{Action}}}
\newcommand{\Intention}{\ensuremath{\mathit{policyPath}}}

%% file: abstract.tex
\begin{abstract}
 Network operators today spend significant manual effort in ensuring and 
checking that the network  meets their intended policies. While recent work in 
 network verification  has made giant strides to reduce this effort, 
  they focus on simple reachability properties and cannot handle {\em
context-dependent} policies (e.g., how many connections has a host spawned)
that operators realize using {\em stateful}  network functions (\DPFs).
 Together, these introduce new expressiveness and scalability challenges that 
 fall outside the scope of existing network verification mechanisms. 
To address these challenges,  we present
 \Name, a system  that enables operators to
test if network with stateful data plane elements correctly implements a given
context-dependent policy.  Our  design makes three key contributions to
address expressiveness and scalability: (1) An abstract I/O unit for  modeling
network I/O  that encodes policy-relevant context information; (2) A practical
representation of complex \DPFs via an  ensemble of finite-state machines
abstraction; and (3) A scalable application of symbolic execution to tackle
state-space explosion.  
We demonstrate that \Name  is several orders of magnitude faster than   existing mechanisms.


 \end{abstract}

%% file: intro.tex
\section{Introduction}
\label{sec:intro}
Network policy enforcement has been and continues to be a challenging and
error-prone task. For instance,  a recent operator survey found that 35\% of
networks generate $\geq$ 100 problem tickets per month and one-fourth of these
take multiple engineer-hours to resolve~\cite{trouble_survey}.  In this
respect, recent efforts on network testing and  verification
(e.g.,~\cite{hsa,veriflow,netplumber, vericon}) offer a promising alternative
to existing expensive and manual debugging efforts.

Despite these advances, there are
fundamental gaps between the intent of network operators and the capabilities
of these tools on two fronts: (1)  data plane elements are
complex and {\em stateful} (e.g., a TCP connection state in a stateful
firewall)  and (2) actual policies  are  {\em context dependent}; e.g.,
compositional requirements to ensure traffic  is ``chained'' through services~\cite{nsx,nfv_2012} or dynamically triggered 
  based on observed host behavior~\cite{intel_on}.

Together, stateful data planes and context-dependent policies introduce new
challenges that  fall outside the scope of existing network
checking mechanisms~\cite{hsa,atpg,veriflow,netplumber}.
 To understand why, it is useful to revisit
their conceptual basis. Essentially, they capture  network
behavior by modeling each 
{\em network function}\footnote{A network function may be stateless 
(i.e., switches/routers) or stateful (i.e., middleboxes) and can be 
physical or virtual.} 
(\DPF) (e.g., a switch) 
as a ``transfer'' function $\Transfer(\header,\port)$ that takes in  a
{\em located packet} (a header $\header$ and a port $\port$) and outputs another
located packet.\footnote{For concreteness, we borrow  terminology from
HSA~\cite{hsa}; other efforts share similar ideas at their
core~\cite{veriflow,netplumber,msr_nod,pandaarxiv}.} Then, some  {\em search
algorithm} (e.g., model checking or geometric analysis) is used to reason about
the composition of these $\Transfer$ functions. Specifically,   we identify
three key limitations with respect to {\em expressiveness} and {\em
scalability} (\Section\ref{sec:motivation}):

\begin{figure}[t]
\centering
\includegraphics[width=190pt]{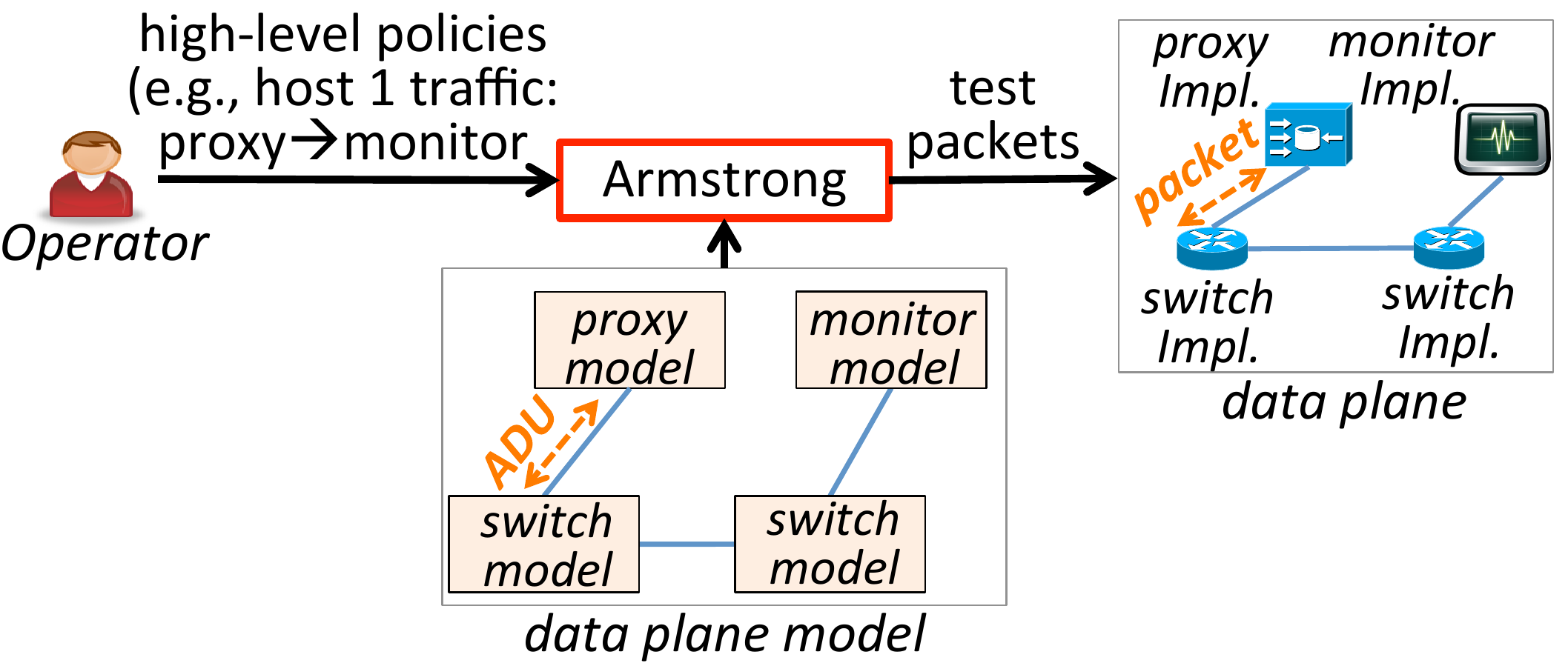}
\tightcaption{ \Name takes in  high-level 
 policy intent from the network operator and  generates test cases to check 
 the implementation of the policy.}
\label{fig:buzz_model_philosophy}
\end{figure}

\begin{packeditemize}

\item {\bf Packets are cumbersome and insufficient:} While  located packets
allow us to compose models of \DPFs, they are inefficient  to
capture higher-layer processing semantics  (e.g., proxy at HTTP level).
Further, in the presence of dynamic middlebox actions~\cite{flowtags_nsdilong},
   located packets  lack
the necessary context information w.r.t.\ a packet's 
 processing history and provenance, which are critical 
 to reason about policies beyond reachability.

\item {\bf Transfer functions  lack state and  context:} The  transfer
abstraction misses key stateful semantics; e.g., reflexive 
 ACLs in a stateful
firewall  or a NAT using consistent public-private IP mappings.
Moreover, the output actions of \DPFs  have richer semantics (e.g., alerts) 
beyond a
located packet  that determine  the policy-relevant context.

\item {\bf Search complexity:}  Exploring data plane behavior is hard even for
reachability properties~\cite{veriflow,netplumber,atpg}.  With stateful
behaviors and richer policies,  exploration is  even more intractable 
 and  existing state-space search algorithms (e.g.,  model checking) can take several tens of hours even on small networks with
$\leq$ 5 stateful \DPFs.

\end{packeditemize}

To address these challenges, we present a network testing framework called {\bf
\Name} (Figure~\ref{fig:buzz_model_philosophy}).  We adopt  active data plane
testing to complement static verification~\cite{atpg,pandaarxiv,vericon},
because it gives concrete assurances about the behavior
``on-the-wire''~\cite{atpg}.  \Name takes in high-level network policies from
the operator, generates and injects test traffic into the data plane, and then
reports if the observed behavior matches the policy intent.  Note that \Name is
not (and does not mandate) a specific policy enforcement system~\cite{pyretic,
kinetic, simplesigcomm}; rather it helps operators to check if the intended policy
is implemented correctly.     

 \Name's design makes three key contributions  to address the 
expressiveness and scalability challenges: 

\begin{packeditemize}

\item {\bf \PADU  I/O abstraction (\S\ref{sec:traffic_unit}):} We propose a new
\Name Data Unit (\PADU) as a common denominator  of traffic processing for
network models.  To improve scalability an \PADU represents an {\em aggregate}
sequence of packets; e.g., a HTTP response \PADU coalesces tens of raw IP
packets.  Furthermore, an \PADU explicitly includes the necessary packet
processing context; e.g., an \PADU that induced an alarm carries this
information going forward.

\item {\bf  FSMs-ensemble model for \DPFs (\S\ref{sec:dpmodel}):}  One might 
be tempted to use a \DPF's code or a finite-state machine (FSM) model
as a \DPF's model, as they can capture stateful behaviors.  
However, these  are  intractable  due to the huge number of states and 
transitions (or code paths).  To ensure a tractable representation,   we model  
complex \DPFs as an {\em ensemble of FSMs}  by decoupling logically 
independent tasks (e.g., client-side vs.\ server-side connection in a \DPF) 
and units of traffic (e.g., different TCP connections).

\item {\bf Optimized symbolic execution workflow (\S\ref{sec:traffic-gen}):} 
 For scalable test generation,  we  decouple it into two stages:
(1) abstract test plan generation at the \PADU granularity using symbolic
execution (\SE) because of its well-known scalability 
properties~\cite{state-exp,acmsymbolicexecution}   and (2) a translation stage
to convert abstract plans  into concrete test traffic. We engineer
domain-specific optimizations (e.g.,  reduce  the number and scope of symbolic
variables) to improve the scalability of \SE in our domain.

\end{packeditemize}

We have written  models for several canonical \DPFs in {\tt C}
 and implement our domain-specific \SE optimizations on top of \klee.  We prototype
\Name as an application over {\tt OpenDayLight}~\cite{opendaylight}. 
 We implement simple  monitoring and test validation mechanisms 
  to localize the
\DPF inducing policy violations~(\S\ref{sec:implementation}).
Our evaluation (\S\ref{sec:eval}) on a real testbed reveals that \Name: (1) can
test hundreds of policy scenarios on networks with hundreds of switches and
stateful \DPFs nodes within two minutes; (2)   dramatically improves test
scalability, providing nearly five orders of magnitude reduction in time for
test traffic generation relative to strawman solutions (e.g., using packets as 
\DPFs models I/O, or using model checking for search); (3) is more expressive 
and scalable than the state of the art; (4) effectively localizes intentional
data/control plane bugs within tens of seconds.

%% file: motivation_vyas.tex
\section{Motivation} 
\label{sec:motivation}

In this section, we use small but realistic network scenarios to highlight the
types of  {\em stateful \DPFs} and {\em context-dependent
policies} used by network operators.  We  also highlight key limitations of
existing network test/verification efforts.  To make the discussion concrete,
we  use the transfer function and located packet abstraction from
HSA~\cite{hsa}/ATPG~\cite{atpg}, where each network  \DPF (e.g., a switch) is a
``transfer'' function $\Transfer(\header,\iport)$ whose  input is a located
packet (a header, port tuple) and outputs another located packet.\footnote{For
brevity, we assume no multicast/broadcast effects.} The behavior of the network
is the composition of such functions; i.e., $\Transfer_n(\ldots
(\Transfer_2(\Transfer_1(\header,\iport))))$.  Our goal here is not to show
the limitations of these specific efforts,  but to highlight why the following 
scenarios fall outside the scope of  this  class of verification techniques
(e.g.,~\cite{anteater,veriflow, netplumber}).

\subsection{Stateful firewalling} 
\label{sec2:firewall}

\begin{figure}[t] \centering
\includegraphics[width=170pt]{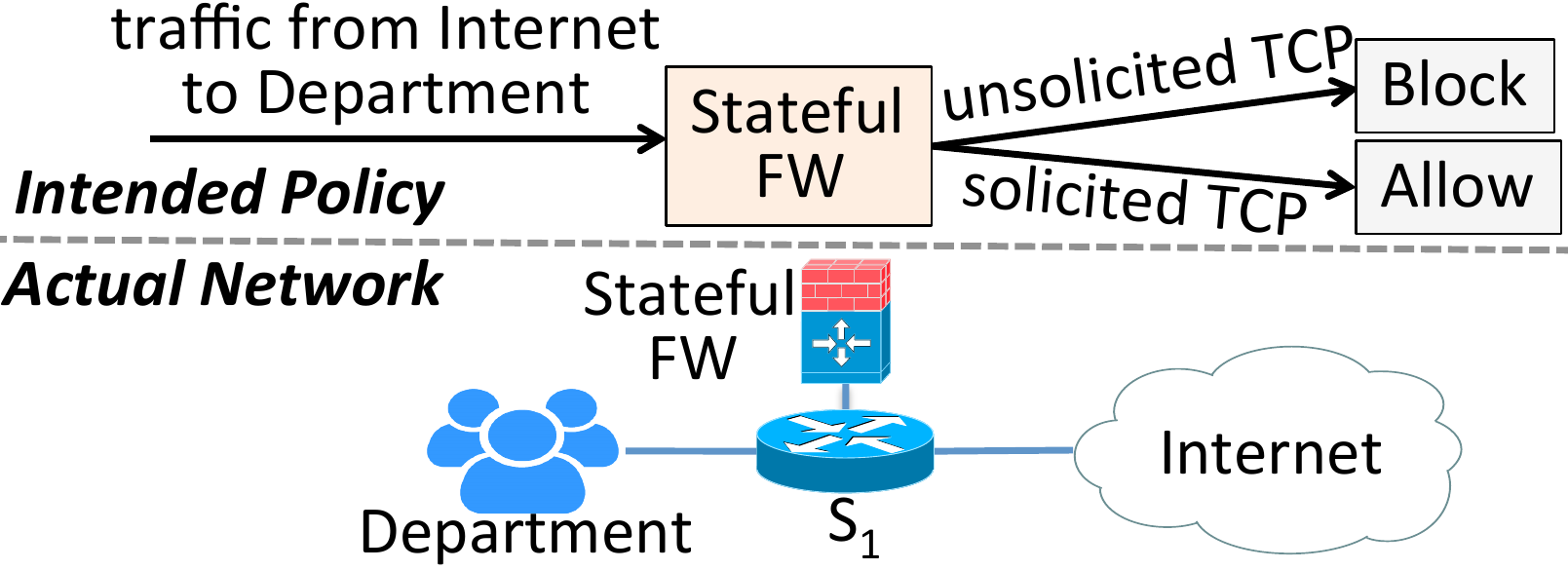}
\tightcaption{Is the firewall allowing solicited and blocking unsolicited traffic from the Internet?} 
\label{fig:one_fw}
\end{figure}

 While simple firewalls and OpenFlow ACLs have a simple match-action operation,
real firewalls capture TCP session semantics.  A common use  is reflexive
ACLs~\cite{reflexsive_acl} shown in Figure~\ref{fig:one_fw},  where the intent
is to only allow  incoming packets for established TCP connections that have
been initiated from ``internal'' hosts.  We depict  the intended policy shown
as a simple policy graph shown on the top.

 Unfortunately, even this simple  policy cannot be captured by a stateless
transfer function $\Transfer(\header,\iport)$.    In particular, the
$\Transfer$ behavior depends on the current {\em state} of the firewall for  a
given connection, and  the function needs to update the relevant internal state
variable.  A natural extension is  a finite-state machine (FSM)  abstraction
where $\Transfer(\header,\iport,\state)$ takes in a located packet and the
current state,  outputs a located packet, and updates the state. In this case,
the state is per-session, but more generally it can span multiple
sessions~\cite{opennf}.


\subsection{Dynamic policy violations} 
\label{sec2:proxy}

\begin{figure}[t] \centering
\includegraphics[width=190pt]{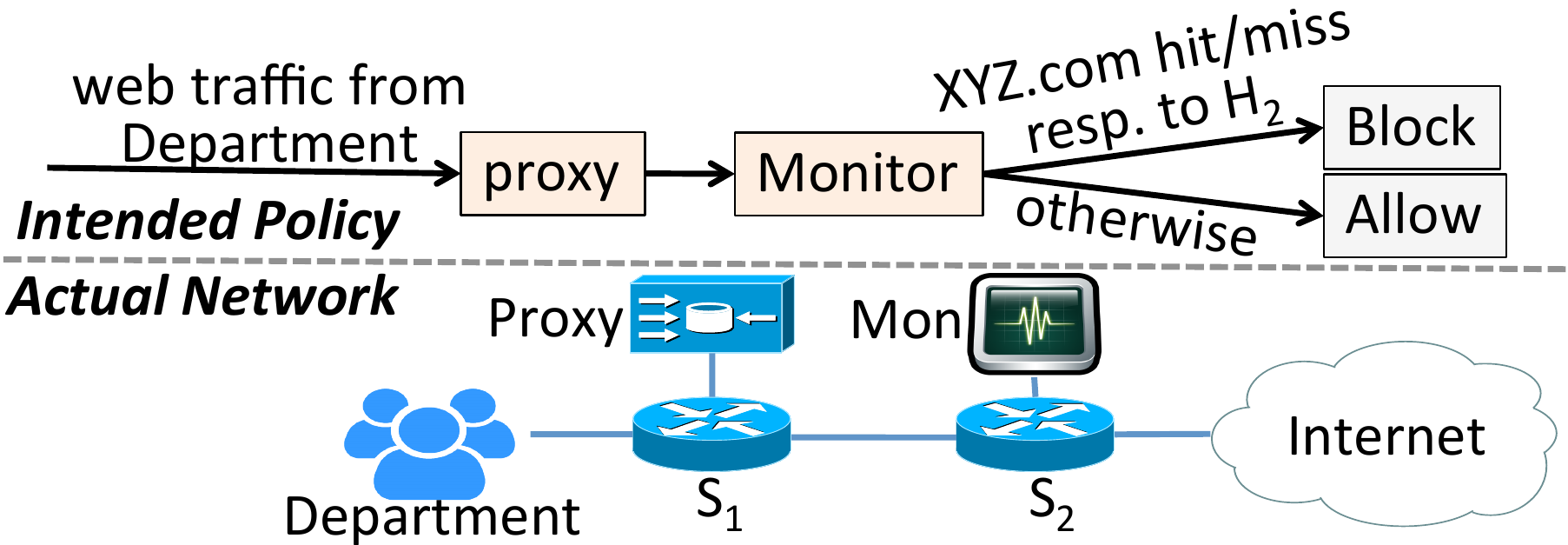}
\tightcaption{Are both hit/miss traffic monitored correctly?} 
\label{fig:proxy_monitoring_policy}
\end{figure}

Next, let us consider  Figure~\ref{fig:proxy_monitoring_policy}, where the
operator  uses a proxy for better performance and also wants to restrict web
access; e.g., $H_2$ cannot access to XYZ.com.  As observed
elsewhere~\cite{flowtags_nsdilong}, there are subtle violations that could occur if a
cached response bypasses the monitoring device.  Prior work has suggested many
candidate fixes; e.g., better \DPF placement, tunnels, or new extended SDN
APIs~\cite{flowtags_nsdilong}.  Our focus here is to check whether such 
policy enforcement mechanisms implement the policy correctly rather than 
developing new  enforcement mechanisms.

As before,  we need to model the stateful behavior of the proxy across connections,
so let us consider our extended  function $\Transfer(\header,\iport,\state)$.
However,   modeling the state alone is not sufficient.  Specifically, the
policy violations happen for  cached  responses, but this \emph{context} (i.e.,
cached or not in this example) depends on some internal state variable 
inside the $\Transfer_\mathit{proxy}$ function.  To faithfully capture the 
policy intent of the operator in our network model, we need to expose 
such relevant  traffic's processing history in our model. This 
suggests that   we need to further extend the functions to include context 
as input  $\Transfer(\header,\iport,\state,\context)$ because  the correct 
network  behavior  (e.g., downstream  switches and middleboxes in our 
model)  depends on this context. We formalized this definitions 
in~\Section\ref{sec:formulation}.

This example also highlights several other issues.  First, different \DPFs operate 
at different layers of the network stack; e.g., the monitoring device may operate 
at  L3/L4 but the proxy in terms of HTTP sessions, which makes  the ``atomic'' granularity  
at which their  policy-relevant states/contexts manifest different.   While it may be tempting
to choose different granularities of traffic for different \DPFs,  it means
 that we may no longer compose our $\Transfer$ functions if their inputs are different. 
Second,  the  policy-relevant context  depends on a {\em sequence} of packets rather 
than on an individual packet.  While it is not incorrect to  think of $\Transfer$ 
functions  operating on packets, it is not an efficient  abstraction.  Finally, note 
that just using  headers is not  sufficient as the behavior of the 
proxy depends on the actual content.

 
\subsection{Firewalling with cascaded NATs} 
\label{sec2:nat}

\begin{figure}[t] \centering
\vspace{0.8cm}
\includegraphics[width=220pt]{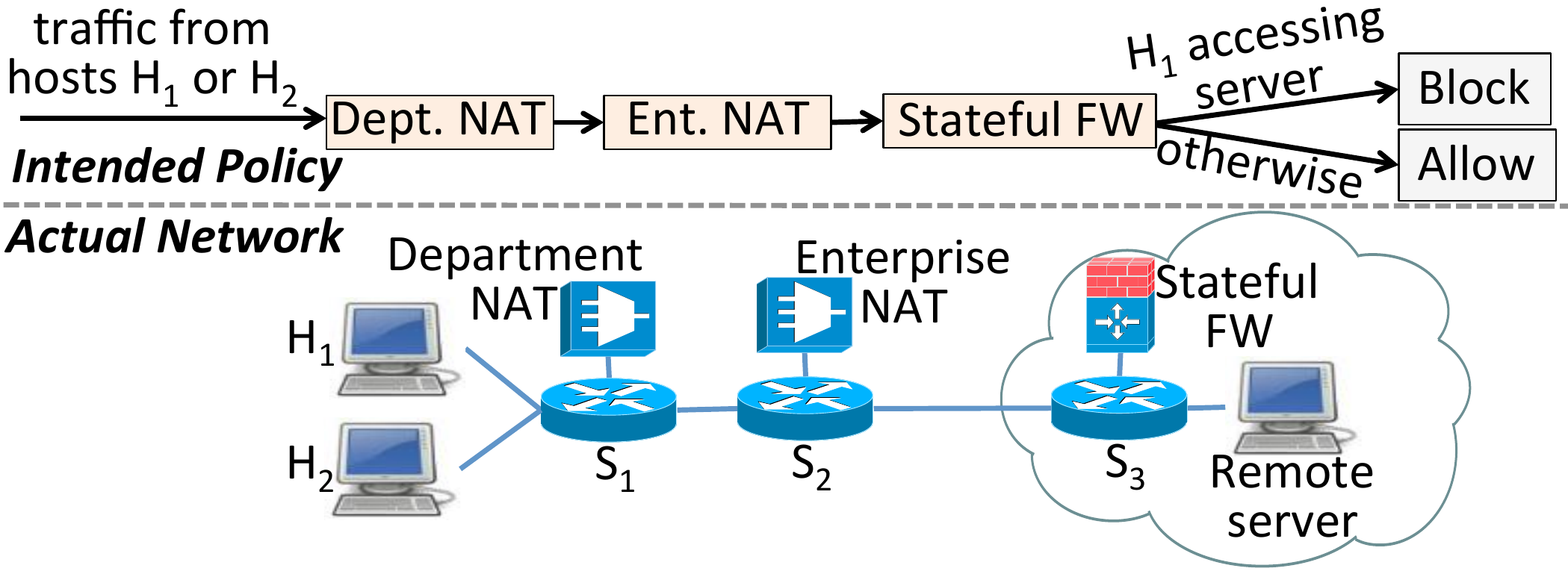}
\vspace{-0.2cm}
\tightcaption{Are we firewalling correctly based on host?} 
\vspace{-0.3cm}
\label{fig:cascaded_nats}
\end{figure}

Figure~\ref{fig:cascaded_nats} depicts a  scenario  inspired by prior work
that showed  cascaded NATs are error-prone~\cite{hotmbox_nats,
netblaster}.  Note that  a correct NAT  should use a  consistent
public-private IP mapping for a session~\cite{symnet}.  
 To model such network behaviors, we need to both  capture the packet 
 provenance (i.e., where it originated from) and  the consistent mapping 
 semantics. 

Unfortunately,  existing tools such as HSA/ATPG  essentially model stateful
\DPFs as ``black box'' functions and do not capture or preserve the  flow consistent
mapping  properties.  This has  two natural implications for our extended
transfer function $\Transfer(\header,\iport,\state,\context)$: (1) the context
$\context$ should  also include the packet provenance, and  (2) the function
$\Transfer$ must be expressive enough to  capture stateful \DPFs semantics 
(e.g., session-consistent mappings).

\begin{figure}[t] \centering
\vspace{0.8cm}
\includegraphics[width=220pt]{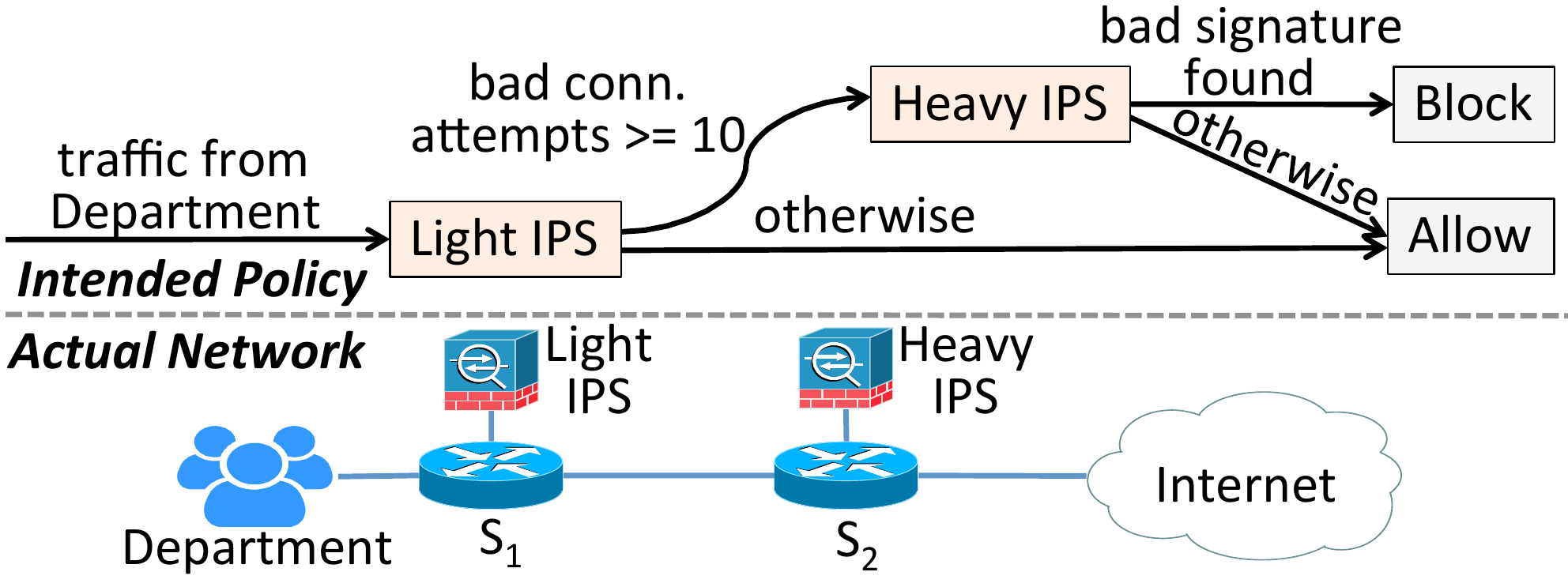}
\vspace{-0.2cm}
\tightcaption{Is suspicious traffic sent to heavy IPS?} 
\label{fig:dynamic_policy}
\end{figure}

\subsection{Multi-stage triggers} 
\label{sec2:ips}

So far our examples underlined  the need for capturing stateful semantics
and  relevant context inside a transfer function. We end
this discussion with a  {\em dynamic service chaining} example in
Figure~\ref{fig:dynamic_policy} that combines both effects.  The intended
policy is to use the light-weight IPS (L-IPS) in the common case (i.e., for
all traffic) and only subject suspicious hosts flagged by the L-IPS 
(e.g., when a host generates too
many scans) to the more expensive H-IPS (e.g., for payload signature matching).
Such multi-stage detection is useful; e.g., to minimize latency and/or reduce
the H-IPS load.   Such scenarios are implemented today (albeit typically 
by hardcoding the policy into the topology) and  enabled by novel SDN-based dynamic
control mechanisms~\cite{kinetic, intel_dynamic_sc}. 
Unfortunately, we cannot check that this multi-stage operation works correctly
using existing reachability mechanisms~\cite{hsa, atpg} because they ignore
the IPSes states (e.g., the current per-host count of bad
connections inside the L-IPS) and traffic context related to the  sequence 
of intended actions.  

Finally, note that the above examples have  natural implications for a
\emph{search strategy} to explore the data plane behavior.  Prior 
exhaustive search strategies were possible only
because a transfer function processes each ``header'' independently and had no
state. Thus they only had to  search over the  ``header space''. Note that this
is already hard and requires clever algorithms~\cite{atpg} and/or parallel
solvers~\cite{libra}.  Designing a search strategy for the examples above is
fundamentally more challenging because we 
 need to consider a bigger ``traffic'' space (i.e., sequences of packets with payloads)
and we need  to efficiently explore a \emph{state space} since processing of a
packet by an \DPF (e.g., a stateful firewall) can change the
behavior of the data plane for future packets.

\subsection{Key observations} 
\label{subsec:strawmans}

 We summarize key expressiveness and scalability
challenges that fall outside the scope of existing network verification
abstractions and search strategies:


\begin{packeditemize}

\item \DPFs are {\em stateful} (e.g.,~\Section\ref{sec2:firewall}) 
and have complex {\em semantics} beyond simple header match-action 
operations, and abstracting them as  blackboxes is insufficient (e.g.,~\Section\ref{sec2:nat});

\item  \DPF  actions are triggered   on  {\em sequences} of packets and occur
at different {\em logical aggregations} (e.g.,~\Section\ref{sec2:proxy});

\item  The correct  behavior depends on traffic  {\em context} 
  such as provenance and processing history 
 (e.g.,~\Section\ref{sec2:proxy} and \Section\ref{sec2:nat});

\item The {\em space}  of possible outcomes in the presence of stateful data
planes operating over richer semantics (e.g., payload)  and context-dependent 
policies can be very large (e.g.,~\Section\ref{sec2:ips}).

\end{packeditemize}

%% file: formulation.tex
\section{Problem Formulation}
\label{sec:formulation}

In this section, 
we define the semantics of a stateful data plane
using which we formalize a test trace to test the intended policies in 
a data plane. We then use these definitions to motivate the need for 
a model-based testing approach.

\subsection{Data Plane Semantics}
\label{subsec:data_plane_semantics}

In this sub-section we formalize the semantics of stateful data 
planes and context-dependent policies. This formalization serves 
two purposes: (1) an understanding of the data plane semantics, 
where actual traffic is processed, provides insight into the methodology of 
generating test traffic; in particular, as we will see in this section, this 
formalization motivates the need for modeling the data plane to 
bridge the gap between a high-level policy and its manifestation in 
the data plane; (2) it serves as a reference point for the future 
research in the area of stateful data planes and context-dependent 
policies\footnote{Previous work has modeled network semantics
without focusing on stateful data planes and context-dependent 
policies (e.g.,~\cite{pyretic, merlin})}.

\mypara{DPF}
Since test traffic operates on the data plane level, in this sub-section
we define the data plane semantics. First, we define the semantics 
of a \DPF and the network.
Let $\packets$ denote the set of packets.\footnote{Packets are
  ``located''~\cite{hsa,consupd}, so that the \DPF can identify and
  use the incoming network interface information in its processing
  logic.}  Formally, a \DPF is a 4-tuple $(\DPFState, \DPFInit,
\Ports, \Transition)$ where: (i) $\DPFState$ is a finite set of
states; (ii) $\DPFInit \in \DPFState$ is the initial state; (iii)
$\Ports$ is the set of network \edges; and (iv) $\delta : \DPFState
\times \packets \mapsto \DPFState \times \packets \times \Ports \times
\ActionSet$ is the transition relation.


Here, $\ActionSet$ is a set of {\em \actions} that capture the
response of a \DPF to a packet.  Each $\ActionVar \in \ActionSet$
provides contextual information that the administrator cares about.
Each $\ActionVar$ is annotated with the specific \DPF generating the
\action and its relevant states; e.g., in
Figure~\ref{fig:dynamic_policy} we can have $\ActionVar_1 = \langle
\mathit{LIPS}: H_1, \mathit{Alarm,SendToHIPS} \rangle$ when the
$\mathit{LIPS}$ raises an alarm and redirects traffic from $H_1$ to
the H-IPS, and $\ActionVar_2 = \langle \mathit{LIPS}: H_1,
\mathit{OK,SendToInternet} \rangle$ when the $\mathit{LIPS}$ decides
that the traffic from $H_1$ was OK to send to the Internet. Using
\actions, administrators can define high level \Policy intents rather
than worry about low-level \DPF states.
%
%
Note that this \DPF definition is general and it
encompasses stateful \DPFs from the previous section and 
 stateless L2-L3 devices.



\mypara{Network} Formally, a network data plane $\netw$ is a pair
$(\nodes, \topo)$ where $\nodes = \set {\DPF_1, \dots, \DPF_\NumDPFs}$
is a set of \DPFs and $\tau$ is the topology map. Informally, if
$\tau(\PortVar) = \DPF_\DPFIndex$ then packets sent out on \edge
$\PortVar$ are received by ${\DPF}_i$.\footnote{We assume each \edge
  is mapped to unique incoming/outgoing physical network ports on two
  different \DPFs.}  We assume that the graph has well-defined sources
(with no incoming \edges), and one more sinks (with no outgoing
\edges). The {\em data plane state} of $\netw$ is a tuple
$\NetState=(\DPFStateVar_1, \dots, \DPFStateVar_\NumDPFs)$, where
$s_i$ is a state of $\DPF_i$.


\mypara{Processing semantics} To simplify the semantics of packet 
processing, we assume packets are
processed in a lock-step (i.e., one-packet-per-\DPF-at-time) fashion
and do not model (a) batching or queuing effects inside the network 
(hence no re-ordering and packet loss); (b) parallel processing effects 
inside \DPFs; and (c) the simultaneous processing of different packets 
across \DPFs.
 
Let $\NetState = (\DPFStateVar_1, \dots, \DPFStateVar_i, \dots,
\DPFStateVar_\NumDPFs)$ and $\NetState' = (\DPFStateVar_1, \dots,
\DPFStateVar'_i, \dots, \DPFStateVar_\NumDPFs)$ be two states of
$\netw$. First, we define a \emph{single-hop} network state transition
from $(\NetState, \DPFIndex, \packet)$ to $(\NetState', \DPFIndex',
\packet')$ labeled by \action $\ActionVar$, denoted $\sstep {
  (\NetState, \DPFIndex, \packet)} {\ActionVar} {(\NetState',
  \DPFIndex', \packet' ) }$ if $\Transition_\DPFIndex (
\DPFStateVar_\DPFIndex, \packet) = (\DPFStateVar'_\DPFIndex, \packet',
\PortVar, \ActionVar)$, with $\DPF_{\DPFIndex'} = \topo(\PortVar)$.  A
\emph{single-hop} network state transition represents processing of
one packet by $\DPF_i$ while the state of all \DPFs other than
$\DPF_i$ remains unchanged. For example, when the L-IPS rejects a
connection from a user, it increments a variable tracking the number
of failed connections. Similarly, when the stateful firewall sees a
new three-way handshake completed, it updates the state for this
session to {\tt connected}.

Next, we define the {\em end-to-end} state transitions that a packet
$\inpacket$ entering the network induces.  Suppose $\inpacket$
traverses a path of length $\PathLength$ through the sequence of \DPFs
$\DPF_{\DPFIndex_1}, \dots, \DPF_{\DPFIndex_\PathLength}$ and ends up
in $\DPF_{\DPFIndex_{\PathLength+1}}$ (note that the sequence of
traversed \DPFs may be different for different packets). Then the
end-to-end transition is a 4-tuple $(\NetState_1, \inpacket,
\seq{\ActionVar_1, \dots, \ActionVar_\PathLength},
\NetState_{\PathLength+1})$ such that there exists a sequence of
packets $\packet_1, \dots, \packet_{\PathLength+1}$ with $\packet_1 =
\inpacket$, and a sequence of network states $\NetState_2, \dots,
\NetState_{\PathLength-1}$ such that $\forall 1 \leq k \leq
\PathLength \such \sstep { ( \NetState_k, \DPFIndex_k, \packet_k ) } {
  \ActionVar_k } { (\NetState_{k+1}, \DPFIndex_{k+1}, \packet_{k+1} )
}$.
 
That is, the injection of packet $\inpacket$ into $\DPF_{\DPFIndex_1}$
when the network is in state $\NetState_1$ causes the sequence of
\actions $\seq{\alpha_1, \dots, \alpha_\PathLength}$ and the network
to move to state $\NetState_{\PathLength+1}$, through the above
intermediate states, while the packet ends up in
$\DPF_{\DPFIndex_{\PathLength+1}}$. For instance, when the L-IPS is
already in the {\tt toomanyconn-1} state for a particular user and the
user sends another connection attempt, then the L-IPS will transition
to the {\tt toomanyconn} state and then the packet will be redirected
to the H-IPS.


Let $\etoesem{\netw}$ denote the end-to-end ``network semantics'' or
the set of feasible transitions on the network $\netw$ for a single
input packet.

\mypara{Trace semantics} Next, we define the semantics of processing
of an input packet trace $\Trace = \inpacket_1, \dots,
\inpacket_\NumPkts$.  We use $\aseq$ to denote the vector of \DPF
\actions associated with this trace; i.e., the set of \actions across
all \DPFs in the network. The network semantics on a trace $\Trace$ is
a sequence of \action vectors: $\TraceSem_\Trace = \langle \aseq_1,
\dots, \aseq_\NumPkts \rangle$ where
$\forall 1 \leq \PktIndex \leq \NumPkts \such \inpacket_\PktIndex \in
\packets \land \aseq_\PktIndex \in {\ActionSet}^+$.  This is an
acceptable sequence of events iff there exists a sequence
$\NetState_1, \dots, \NetState_{m+1}$ of states of $\netw$ such that:
$\forall 1 \leq \PktIndex \leq \NumPkts \such
(\NetState_\PktIndex, \inpacket_\PktIndex, \aseq_\PktIndex,
\NetState_{\PktIndex+1}) \in \etoesem{\netw}$.

\mypara{Policies in the data plane} Given the notion of trace 
semantics defined above, we can now formally
specify our goal in developing \Name.  At a high-level, we want to
test a {\em \intent}. Formally, a \intent is a pair $(\TracePredicate
; \TraceSem)$, where $\TracePredicate$ captures a class of traffic of
interest, and $\TraceSem$ is the vector of \actions of the form
$\langle \aseq_1 \dots \aseq_\NumPkts \rangle$ that we want to observe
from a correct network when injected with traffic from that class.
Concretely, consider two \intents:
\begin{packedenumerate} 
\item In Figure~\ref{fig:proxy_monitoring_policy}, we want: \emph{``Cached
  web responses to Dept1 should go to the monitor''}. Then,
  $\TracePredicate$ captures web traffic to/from Dept1 and $\TraceSem
  = \seq{\alpha_1, \alpha_2}$, with $\alpha_1 = \mathit{Proxy: Dept1,
    CachedObject}$ and $\alpha_2 = \mathit{Proxy: Dept1, SendToMon}$.

\item In Figure~\ref{fig:dynamic_policy} we want: \emph{``If host
  $H_1$ contacts more than $10$ distinct destinations, then its
  traffic is sent to $\mathit{H-IPS}$''}. Then, $\TracePredicate$
  captures traffic from $H_1$, and $\TraceSem = \seq{\alpha_1,
    \alpha_2}$ where $\alpha_1 = \mathit{L-IPS:}
  H_1,\mathit{Morethan10Scan}$, and $\alpha_2 = \mathit{L-IPS}:
  H_1,\mathit{SendtoHIPS}$.

\end{packedenumerate}

\mypara{Test trace generation} Our goal is to check whether such a \intent 
is satisfied by the \emph{actual} network. More specifically, if we have a 
{\em concrete test trace} $\Trace$ that satisfies $\TracePredicate_\Trace$ 
and should {\em ideally} induce the \actions $\TraceSem_\Trace$, then 
the network should exhibit $\TraceSem_\Trace$ when $\Trace$ is injected
into it. In other words, the goal of \Name in terms of test traffic generation 
is to find a concrete trace that satisfies $\TracePredicate_\Trace$.

\subsection{Challenges of automatic test traffic generation}
The vision of \Name involves automating this test traffic
(i.e., a set of test traces corresponding to all policies) generation.
In an attempt to do so, however, we are facing two challenges.
First, operators often define policies using a high-level 
representation, similar to what we saw in~\ref{sec:motivation}, 
as opposed to the complex form $(\TracePredicate; \TraceSem)$ 
that involves low-level intricacies of each \DPF (i.e., 
$(\DPFState, \DPFInit, \Ports, \Transition)$). The challenge of 
test traffic generation is as follows. Given a  policy, how to find 
concrete test traffic, out of very many possible 
distinct traces, that satisfies $\TracePredicate_\Trace$.

In the next two sections we will discuss how \Name overcomes these 
challenges: (1) \Section\ref{sec:dpmodel} will discuss how to \DPF 
models are used to bridge the gap between high-level policies and
low-level data plane semantics; (2) \Section\ref{sec:traffic-gen} then 
will show how to systematically conduct search on the data plane 
model using symbolic execution to generate test traffic.

%% file: overview_vyas.tex
\section{\Name  Overview}
\label{sec:overview}

 In this section, we give an  overview of \Name describing the key components
and design ideas to address the challenges described at the end of the previous
section. 

\mypara{Problem scope}  \Name's  goal is to check if an operator's
intended policy is implemented correctly in the data plane.  
(\Name does not mandate a specific control- or data-plane policy 
enforcement 
mechanism~\cite{vericon,kinetic,simplesigcomm,flowtags_nsdilong,pyretic}, 
and our focus in this work is not on designing such a mechanism.) In this respect,
there are two complementary classes of approaches: (1) {\em Static
verification} (e.g., HSA~\cite{hsa}, Veriflow~\cite{veriflow},
Vericon~\cite{vericon}) in which a model of  the network is given to a
verification engine that checks if the configuration meets the policy (or
produces a counterexample); and (2)  {\em Active testing} (e.g.,
ATPG~\cite{atpg}), where test traffic is injected into the network and check if
the  observed behavior is consistent with the intended policy.
From a practical view,  active testing can detect implementation
problems that is  outside the scope of static verification;  a  bug in the
firewall implementation or the middlebox orchestration 
logic~\cite{splitmergelong, opennf}. Thus, we  adopt an {\em active testing} 
approach in \Name.  That said,  our modeling contributions  will also improve 
the scalability of  static verification.  

\mypara{Scope of policies} For concreteness, we scope  policies
that \Name can (and cannot) check.  In \Name, a \emph{policy} 
is defined as a set of  \emph{policy scenarios}. A policy scenario 
is a 3-tuple  $(\TracePredicate;\Intention;\Action)$.  $\TracePredicate$ 
specifies the traffic class (e.g., in terms of 5-tuple) to which the 
policy is related (e.g., srcIP$\in$Dept, proto=TCP, and dstPort=80
in Figure~\ref{fig:proxy_monitoring_policy}),  $\Intention$ is the 
intended sequence of stateful \DPFs that the traffic needs to go 
through along with the relevant context (e.g., provenance=$H_2$ and 
proxyContext=$<$hit,XYZ.com$>$) and $\Action$ is the intended final 
action (e.g., Drop) on any traffic that matches $\TracePredicate$ 
and $\Intention$. The intended policy of 
Figure~\ref{fig:proxy_monitoring_policy} captures three such 
different possibilities for the intended behavior, namely, 
one ending in action Allow, and two (i.e., for hit and miss) 
ending in action Drop when $H_2$ tries to get XYZ.com, so
the intended policy corresponds to three policy scenarios.

Other properties like checking performance, crash-freedom, infinite loops
inside \DPFs, and race condtions are outside the scope of \Name.  Similarly, if
there are  context/state behaviors  outside the \Name models,
then \Name will not detect those violations.  



\vyas{can u decouple context and chain?}

\mypara{Design space and strawman solutions} Given the complexity of stateful
\DPFs and context-dependent policies,  it will be  tedious for an operator to
manually reason about their interactions and generate concrete test cases to
check the data plane behavior.  In a nutshell, the goal of \Name is  to
simplify the operators workflow so that they only need  to specify high-level
policy scenarios such as the policies from the previous section. \Name
automatically generates test traffic  to exercise  each given policy scenario 
to simplify the process of validating if the data plane correctly implements 
the operator's intention.

In a broader context, \Name is an instance of a specification-based or
model-based testing paradigm~\cite{mbt}. Any model-based testing solution
needs a to bridge the semantic  gap between the high-level intended 
behavior of the system (in case of \Name, high-level policies and the actual 
system behavior (in case of \Name, running code and hardware in the 
data plane).  A specific solution can be viewed in terms of a design 
space involving three key components: (1)  A basic unit of {\em input-output} 
(I/O) behavior; (2) A \emph{model} of the expected behavior of each 
component;  and (3)  Some way to {\em search} the space of end-to-end 
system behaviors to generate test cases.   We, therefore, can represent 
a point from the design space as a 3-tuple with specific designs for each 
component.

 To see why it is challenging to find a solution that is both {\em expressive}
and {\em scalable}, let us consider two  points from this design space. At the 
one end of the spectrum, we have prior work like ATPG~\cite{atpg} with:
$\langle\mathit{I/O=LocPkt},\mathit{Model=Stateless},\mathit{Search=Geometric}\rangle$.
As argued earlier, these are not expressive.  At the opposite end we  
consider running  model checking on implementation source code and 
use packets as the I/O unit; i.e.,
$\langle\mathit{I/O=Pkt},\mathit{Model=Impl},\mathit{Search=MC}\rangle$. While
this can be expressive (modulo the hidden contexts),  it is not scalable given
that actual \DPF code can be tens of thousands of lines of code   
since model checking tools struggle beyond a few hundred lines of code.  
Furthermore, using a \DPF implementation code as its model is problematic, 
as implementation bugs can defeat the purpose of testing by affecting the 
correctness of test cases.\footnote{There is also the pragmatic issue that we 
may not have the actual code for proprietary \DPFs.}

\comment{
\begin{figure}[t] \centering
\includegraphics[width=190pt]{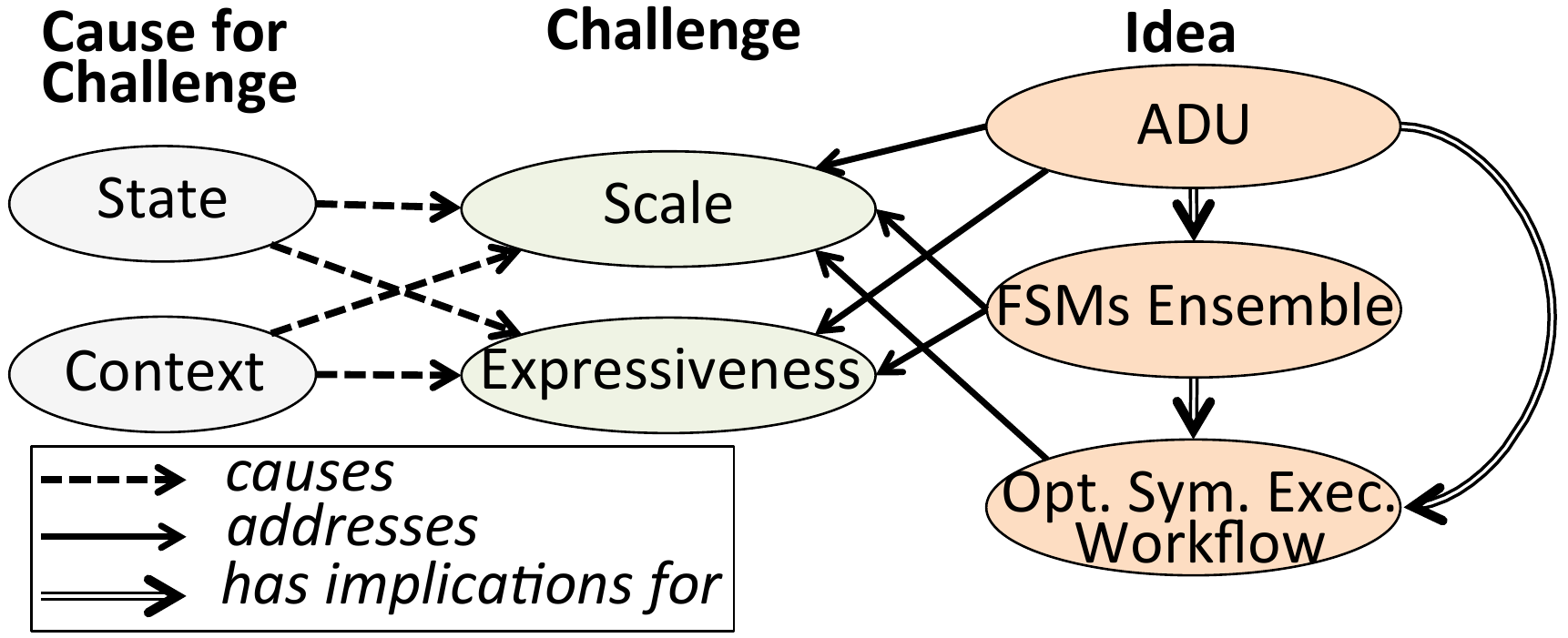}
\vspace{-0.2cm}
\tightcaption{\Name challenges and key ideas.} 
\vspace{-0.2cm}
\label{fig:ideas}
\end{figure}
}

\mypara{High-level approach} Our contribution lies in design choices for each of 
these three dimensions that combine to achieve  scalability and expressiveness: 

\begin{packeditemize}

\item {\bf $\mathit{I/O=\PADU}$ (\S\ref{sec:traffic_unit}):} We  introduce  a novel
abstract network data unit called an \Name Data Unit (\PADU) that improves
scalability of test traffic generation via traffic aggregation and addresses
expressiveness by explicitly capturing relevant traffic context;

\item {\bf $\mathit{Model=FSM Ensemble}$ (\S\ref{sec:dpmodel}):} We model
each \DPF as an ensemble of FSMs and compose them to model the data 
plane.  Here, using FSMs as building blocks enables the stateful model and 
breaking a monolithic FSM into the  ensemble dramatically shrinks the state 
space;

\item {\bf $\mathit{Search=Optimized}$ $\mathit{Symbolic}$ $\mathit{Execution}$
(\Section\ref{sec:traffic-gen}):} Given our goal is to generate test traffic, 
we  can sacrifice exhaustive searching and use more scalable approaches 
like   symbolic execution (\SE)  rather than model-checking. However, 
using \SE naively does not handle large topologies and thus we implement 
domain-specific optimizations for pruning the search space.
\end{packeditemize}

Note that these decisions have natural synergies; e.g., \PADUs simplify the
effort to write \DPF models and also improves the scalability of our \SE
step. 

\begin{figure}[t] \centering
\includegraphics[width=190pt]{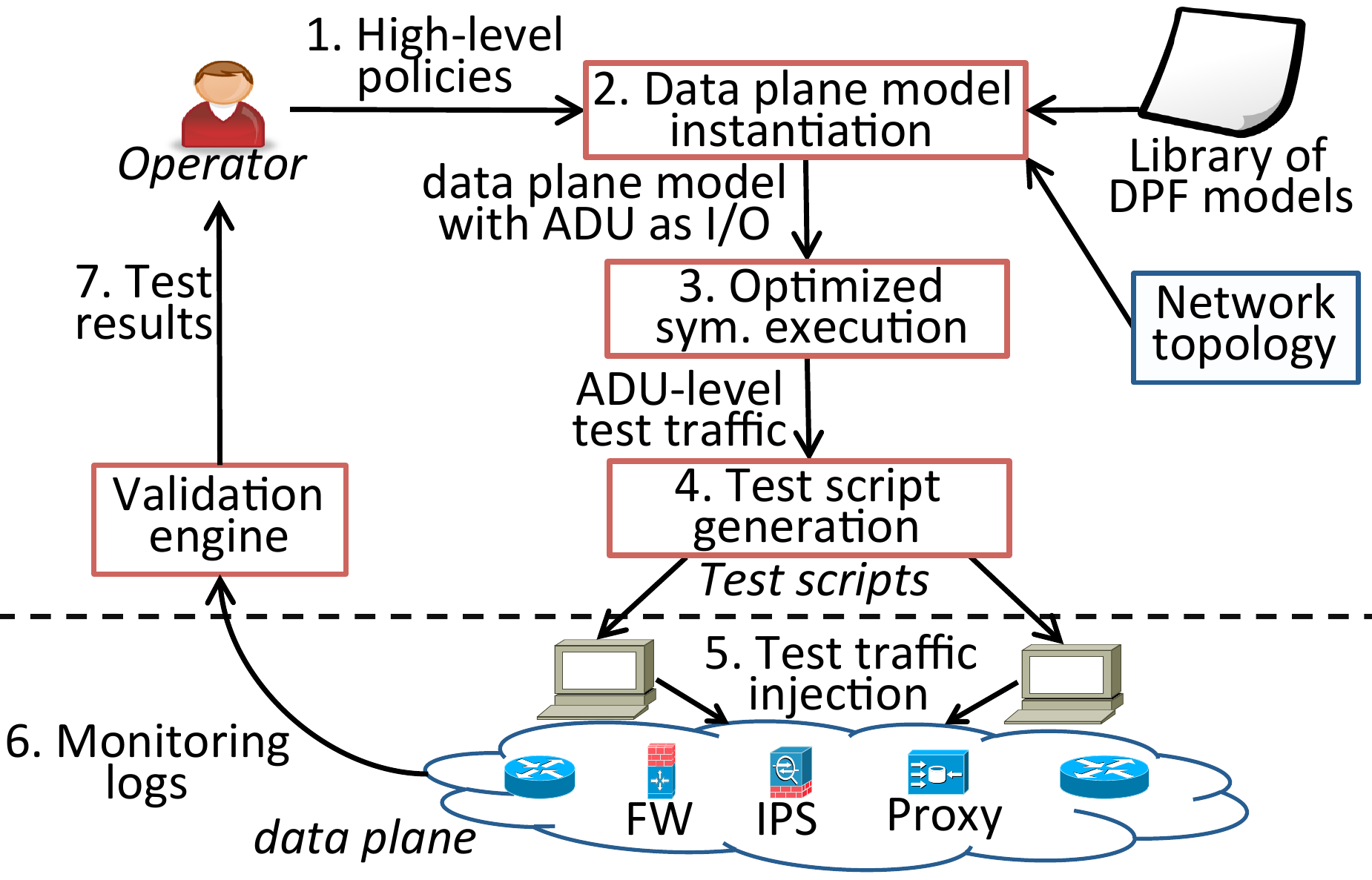}
\vspace{-0.2cm}
\tightcaption{\Name workflow.} 
\label{fig:workflow}
\end{figure}

\mypara{End-to-end workflow} Putting these ideas together,
Figure~\ref{fig:workflow} shows \Name's  end-to-end workflow:
  The operator defines the intended network policies in a 
  high-level form, such as the policy graphs shown on top of 
  each figure of~\S\ref{sec:motivation}. \Name uses a library of \DPF models,
where each model works at the  \PADU granularity. Given the library of \DPF
models and the network topology specification (with various switches and
middleboxes),   \Name constructs a concrete network model for the given
network. Then, it  uses the network model in conjunction with the policies 
to automatically generate concrete  test traffic.   Here,  we decouple the test
traffic generation into two logical steps by first running  \SE on the data plane
model to generate abstract (i.e., \PADU-level) test traffic and then using a  
suite of traffic injection libraries to translate this abstract traffic into concrete  
test traffic via test scripts.  Finally, we use a
monitoring mechanism that records  data plane events and  analyzes 
them to declare a test verdict to the operator (i.e., success, or a policy 
violation along with the \DPF in charge of the violation).

Note that operators do not need to be involved in the task of writing \DPF
models  or in populating  the test generation library.  These are one-time
offline tasks and can be augmented with community efforts~\cite{irtfnfv}.


%% file: traffic_unit.tex
\section{\PADU Input-Output Abstraction}
\label{sec:traffic_unit}

In this section, we present our \PADU abstraction for modeling \DPF I/O
operations and show how it enables scalability and expressiveness, while still
acting as a common denominator across diverse \DPFs.  We  discuss the
implications of this choice for the design of the \DPF models   and our search
strategy.  We end the section with guidelines and a recipe for extending \PADUs
for future scenarios.

\begin{figure}[t]
\vspace{0.1cm}
\centering
\begin{minipage}[b]{200pt}
\begin{lstlisting}[numbers=left,numberstyle=\tiny,frame=single,escapechar=',caption=\PADU structure.\label{lst:padustruct}]
 struct '\PADU'{
  // IP fields 
  int srcIP, dstIP, proto; 
  // transport
  int srcPort,  dstPort;
  // TCP specific 
  int tcpSYN, tcpACK, tcpFIN, tcpRST;
  // HTTP specific 
  int  httpGetObj, httpRespObj; 
  // '\Name'-specific 
  int dropped, networkPort, '\PADU'id; 
  // Each '\DPF' updates traffic context
  int '\actionTag'[MAXTAG];
  ...
 };
\end{lstlisting}
\end{minipage}
\vspace{-0.8cm}
\end{figure}

\mypara{Key ideas} Concretely, an \PADU is simply a {\em struct} as shown in
Listing~\ref{lst:padustruct}.   Our \PADU  abstraction extends  located packets
from prior work in two key ways:

\begin{packeditemize}
 \item {\em Traffic aggregation:} First,  each \PADU can represent a {\em sequence}
of packets rather than an individual packet.  This enables us to 
 represent higher-layer operations more efficiently; e.g., state inside an 
 \DPF (e.g., a TCP connection's current state on a firewall) is associated with a set of packets
rather than a single IP packet. As another example, a proxy's cache state
transitions to a new ``relevant state'' (i.e., cached state with respect to an
object) only after the entire payload has been reassembled.

 \item {\em Explicitly binding the context:} Each \PADU is explicitly bound to its
relevant context through the {\tt \actionTags} field.  Conceptually, {\tt
\actionTags} ensure that the \PADU carries its ``policy-related processing
history'' as it goes through the network.  The natural question is  what should
these  {\tt \actionTags} capture?  Building on our insights from the motivating
examples of~\Section\ref{sec:motivation}, {\tt  \actionTags} contain
two types of information: (1) \PADU's
provenance (i.e., it's origin that may be otherwise hidden, for example, after
a NAT), (2) \DPF processing context for the intended policies (e.g., 1 bit for
cache hit/miss, 1 bit for alarm/no-alarm).  Concretely,  a {\tt \actionTag} is
the union of different fields to embed relevant context w.r.t.\ different
\DPFs that the  \PADU  has gone through and the \PADU provenance. 

\end{packeditemize}

\mypara{Implications for \DPF models and test generation}
The \PADU abstraction has natural
synergies and implications for both \DPF models and test traffic
generation.  First, \PADUs help reduce the complexity of a \DPF's models by
consolidating protocol semantics (e.g., HTTP, TCP) and \actions involving
multiple IP packets.  For example,  all packets corresponding to an HTTP reply
are represented by one \PADU with the {\tt httpRespObj} field indicating the
retrieved object id. Note in particular that the struct fields are a superset
of required fields of specific \DPFs; each \DPF processes only fields relevant
to its function (e.g., the switch function ignores HTTP layer fields of input
\PADUs---see~\Section\ref{sec:dpmodel}.) Second, w.r.t.\ our test traffic
generation,  by aggregating multiple packets, \PADUs reduce the search 
space for model exploration tools such as \SE (\Section\ref{sec:traffic-gen}). 
That said, they introduce a level of indirection because the output of \SE 
cannot be directly used as a test trace and thus we need the extra 
translation step before we can generate  raw packet streams.


\mypara{Designing future \PADUs} Given the continued evolution of \DPFs and
policies, a natural question is how can we extend the basic \PADU.  While we
cannot claim to have  an \PADU definition that can encompass all possible
network scenarios and  policy requirements, we present  a high-level design
roadmap that has served us well. First, the key to determining the fields of an
\PADU is to identify policy-related network protocols in all \DPFs of interest.
For example, each of {\tt TCP SYN}, {\tt TCP SYN+ACK}, etc. make important
state transitions in a stateful firewall and thus should be captured as \PADU
fields. The key point here is that our  \PADU abstraction is future-proof; e.g.,
if we decide to add an ICMP field to the \PADU of Listing~\ref{lst:padustruct}
(e.g., because our new policy involves ICMP on some new \DPF models), this is
not going to affect existing \DPF models, as they simply ignore this new
field. The second point is to consider a conservatively large {\tt \actionTag}
field to accommodate various types of relevant traffic context (e.g., sufficient
number of bits to allow representation of different types of IPS alarms, as 
opposed to having 1 bit for capturing alarm/no-alarm  in {\tt \actionTag}).

%% file: dpmodels.tex
\section{Modeling the Data Plane}
\label{sec:dpmodel}
\comment{
\seyed{
TODO: (undecided to say this here or implementation): make it clear 
that each \DPF model has a ``skeleton'' that reflects its behavior, and 
a set of ``parameters'' that get initialized given high-level policies 
(e.g., the count threshold of L-IPS could be 10). the same way variables
are instantiated by values in programs...}}

 In this section,  we begin by exploring some seemingly natural  strawman
 approaches to  model each \DPF (\Section\ref{subsec:strwaman_models}) and
then present our idea of modeling  \DPFs as ensembles of FSMs by decoupling an
\DPF's actions based on  logically independent units of traffic and internal 
tasks (\S\ref{subsec:ensembles}).


\subsection{Strawman solutions}
\label{subsec:strwaman_models}

To serve as a usable basis for automatic test traffic generation, a \DPF model needs to be
scalable, expressive, and amenable to composition for network-wide modeling.
 Given the composability requirement, we first rule out very ``high-level'' models such 
 as   writing a proxy  in terms of HTTP object requests/responses~\cite{flowtest_hotsdn}. 
 This leaves us two options: using the code or the FSM abstraction we alluded to
 \Section\ref{sec:motivation}.

\begin{packedenumerate}

\item \emph{Code as  ``model''}: This choice seems to remove the burden of
explicit modeling, but such a model is too complex. For instance, 
Squid~\cite{squid} has $\geq$ 200K lines of code and   introduces other sources 
of complexity that are irrelevant to the policies being checked. Another fundamental 
issue with this choice would be that a bug in the to-be-tested   implementation code
affects the correctness of test traffic generated from such model!
 In summary, this approach yields expressive 
 ``models'', but is not scalable for exploring the search space. 

\item \emph{Write an \DPF model as a monolithic FSM:}
\Section\ref{sec:motivation} already suggests that FSMs may be a natural
extension to the stateless transfer functions. Thus, we can consider
each \DPF as an FSM operating at the \PADU granularity.  That is  we can think
of the current state of a stateful \DPF as vector of state variables (e.g., in
proxy this vector may have three elements: per-host connection state,
per-server connection state, and per-object cache state).  Again,  this is
 not scalable; e.g.,  a stateful
\DPF with  $S$ types of state  with $V$ possible values, means 
this ``giant'' FSM  has  $V^S$ states.

\comment
{
\item \emph{High-level models using  ``relevant'' states and inputs}: An
example of this approach is writing a proxy model   in terms of HTTP object
requests/responses~\cite{flowtest_hotsdn}. Given the diversity of \DPF
operations that act on different network stack layers,   such models are fundamentally
non-composable as the input-output granularity of different \DPFs will not
match; e.g., we cannot   simply ``chain'' the output of a proxy operating at
this level to a packet-level firewall. This high-level model is expressive and 
 scalable (only captures relevant states), but is not composable.
} 

\end{packedenumerate}

Based on this discussion, we adopt FSMs as a natural starting point to avoid
the logical problems associated with using code. Next we discuss how we address
the  scalability challenge.

\subsection{Tractable models via \fsmensembles}
\label{subsec:ensembles}

Our insight is to borrow from the design of actual \DPFs.  In practice, 
\DPF programs (e.g., a firewall) do not explicitly enumerate the full-blown 
FSM. Rather, they  independently track the states  for 
 ``active'' connections. Furthermore, different
functional components of an \DPF are naturally segmented; e.g., 
 client- vs.\ server-side handling in a proxy is separate.
 This enables us to simplify a monolithic  \DPF 
FSM into a more tractable {\em ensemble of FSMs} along two dimensions:

\begin{figure}[t]
\centering
\subfloat[A monolithic FSM model of a stateful firewall w.r.t two TCP connections.]
{
\vspace{-0.6cm}
\includegraphics[width=120pt]{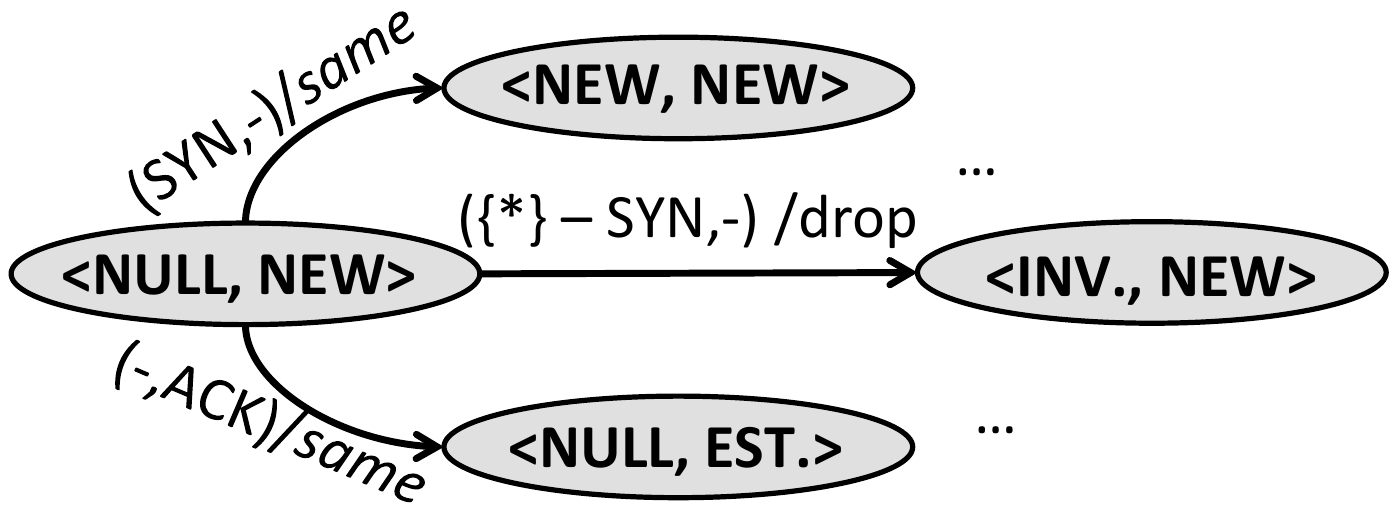}
\label{fig:bad_fw_fsm}
}
\hspace{0.1cm}
\subfloat[A per-connection FSM model of a stateful firewall.]
{
\vspace{-0.6cm}
\includegraphics[width=100pt]{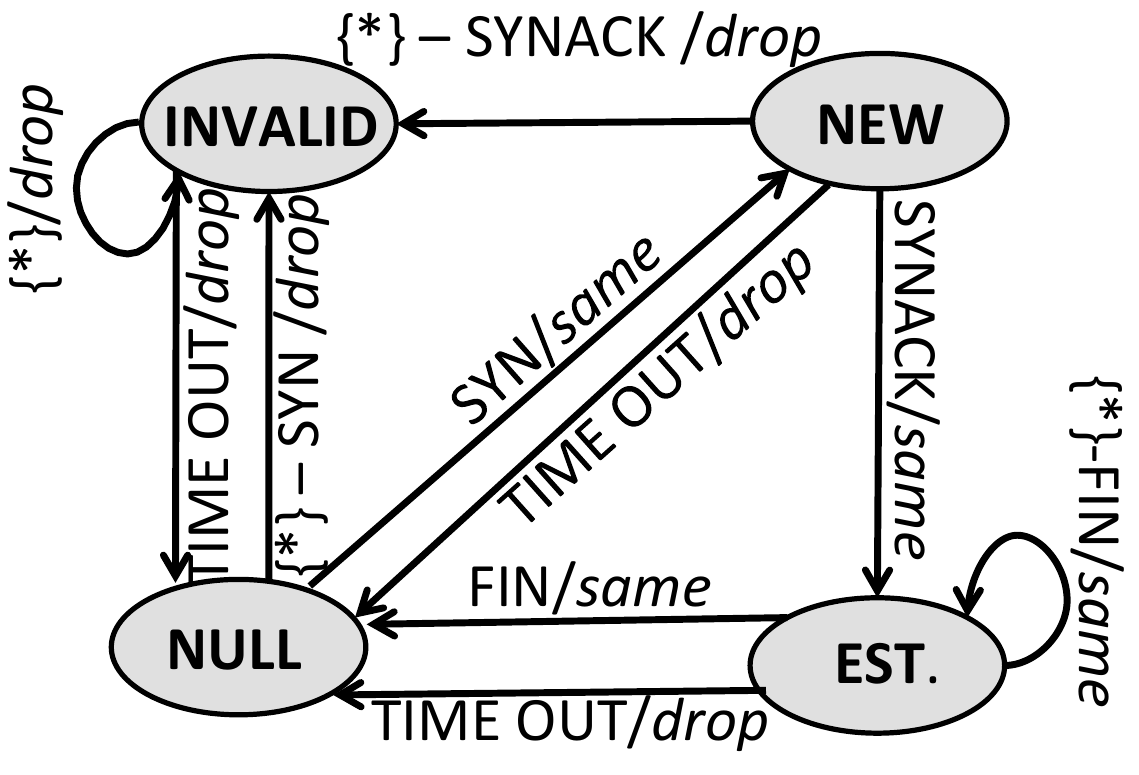}
\label{fig:good_fw_fsm}
}
\tightcaption{Illustrating how decoupling independent traffic units reduces number of 
states.}
\label{fig:fw_fsm}
\end{figure}

\begin{figure}[t]
\centering
\subfloat[A monolithic FSM model of a proxy w.r.t. a client, a server, and an object.]
{
\vspace{-0.6cm}
\includegraphics[width=115pt]{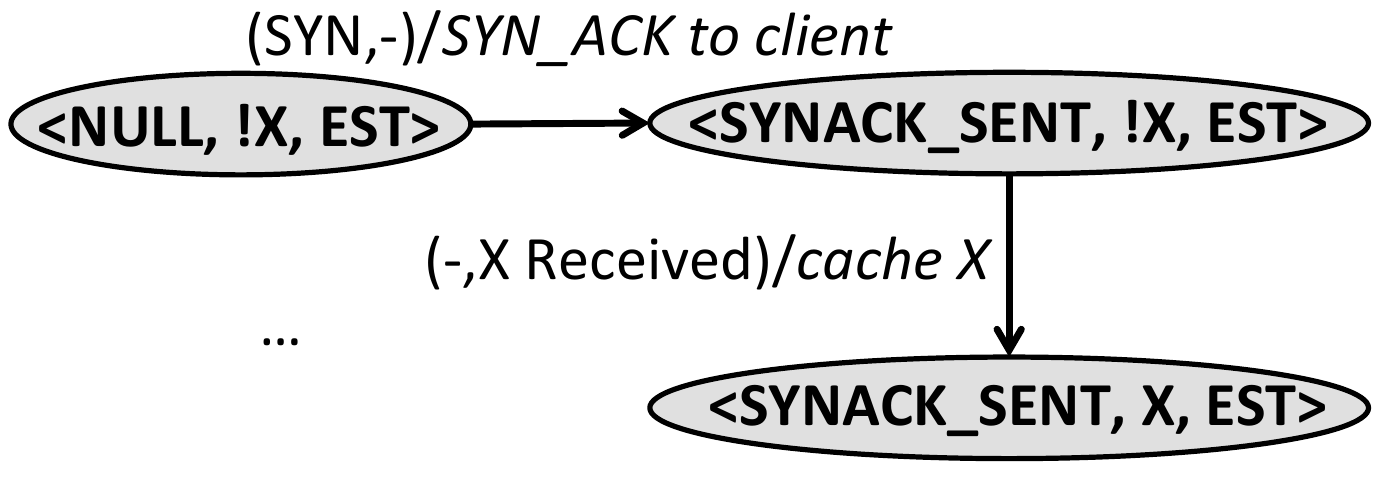}
\label{fig:bad_proxy_fsm}
}
\hspace{0.1cm}
\subfloat[A proxy model as an FSM ensemble, enabled by decoupling independent \DPF tasks.]
{
\vspace{-0.6cm}
\includegraphics[width=105pt]{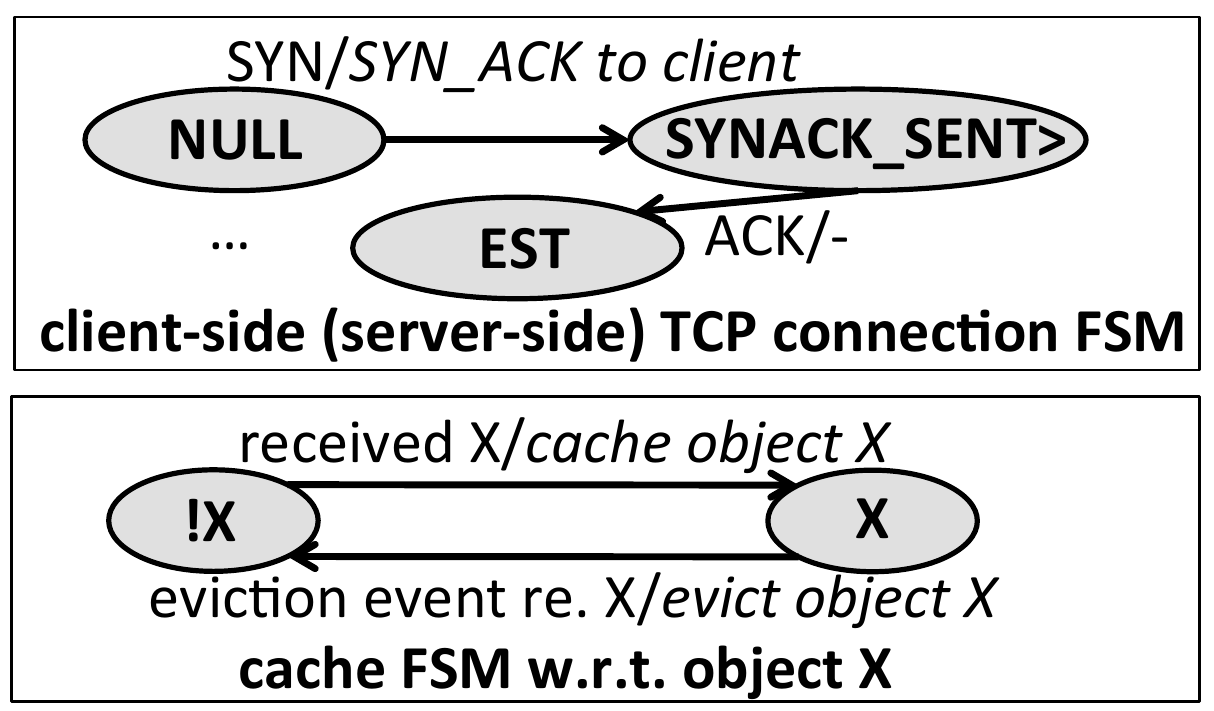}
\label{fig:good_proxy_fsm}
}
\tightcaption{Illustrating how decoupling independent \DPF tasks reduces 
 number of states.}
\label{fig:proxy_fsm}
\end{figure}

\begin{packeditemize}

\item {\bf Decoupling independent traffic units:} Consider a stateful firewall.
The naive approach in modeling it as a monolithic FSM is shown in
Figure~\ref{fig:bad_fw_fsm}.  While this is an expressive model, it is not
scalable as the number of connections grow.  We  decouple this into independent
per-connection FSMs as shown in Figure~\ref{fig:good_fw_fsm}, yielding and the
firewall model as an ensemble of FSMs.\footnote{In general, if the number of
connections is $|conn|$ (2 in this example) and the number of states per
connection is $|state|$ (4 in this example), it is easy to see that this
insight cuts the number of states from $|state|^{|conn|}$ to
$|conn|\times{|state|}$.}

\item {\bf Decoupling  independent tasks:} To see this idea, consider a proxy 
which is instructive, as it operates on a session layer, terminates 
sessions, and it can respond directly with objects in its
cache. The code of a proxy, e.g., {\tt Squid}, effectively has three
 modules: TCP connections with the client, TCP connection with the
server, and cache.  The proxy FSM is effectively the ``product'' of these
modules (Figure~\ref{fig:bad_proxy_fsm}).
However, we can decouple different tasks; i.e.,
client-, server-side TCP connections, and cache.  Instead of a ``giant''
FSM model with each state being of the ``cross-product'' form $\langle
\mathit{client\_TCP\_state,server\_TCP\_state, cache\_content}\rangle$, we use a
 {\em ensemble} of three small FSMs each with a single type of state,
i.e., $\langle\mathit{client\_TCP\_state}\rangle$, $\langle\mathit{server\_TCP\_state}\rangle$, and $\langle \mathit{cache\_content}
\rangle$ (Figure~\ref{fig:good_proxy_fsm}).\footnote{Concretely, if an \DPF has $|T|$ independent tasks  (e.g., 3
for proxy) where the $i$th task has $S_i$ states (e.g., 2 for the cache task in
this example) the ensemble  cuts down the number of states from
$\prod_{i=1}^{|T|} |S_i|$ to $\sum_{i=1}^{|T|} |S_i|$.}

\end{packeditemize}

Note that these ideas are complementary and can be  combined to reduce the
number of states.  For instance, if our proxy is serving two clients talking
to two separate servers, we can first decouple states at the task-level and
further decouple the states within each task at the connection level.

To see this concretely, Listing~\ref{lst:proxyfsmens} shows a partial code snippet 
of a proxy model, focusing on the actions when a client is requesting a
non-cached HTTP object and the proxy does not currently have a TCP
connection established with the server. Here the {\tt id} allows us to
identify the specific proxy instance.  The specific state variables of
different proxy instances are inherently partitioned per \DPF instance
(not shown).  These track the relevant \DPF states, and are updated by
the \DPF-specific functions such as
\texttt{srvConnEstablished}.\footnote{This choice of passing ``id''s
  and modeling the state in per-id global variables is an
  implementation artifact of using C/\klee, and is not fundamental to
  our design.}
If the input {\tt in\PADU} is a client HTTP request
(Line~\ref{lst:proxyfsmensem:req}), and if the requested object is not
cached (Line~\ref{lst:proxyfsmensem:notcache}), the proxy checks the
status of TCP connection with the server.  If there is an existing TCP
connection with the server (Line~\ref{lst:proxyfsmensem:hasconn}), the
output \PADU will be a HTTP request
(Line~\ref{lst:proxyfsmensem:httpreq}). Otherwise, the proxy will
initiate a TCP connection with the server
(Line~\ref{lst:proxyfsmensem:newconn}).

\begin{figure}[t]
\centering
\begin{minipage}[b]{220pt}
\begin{lstlisting}[numbers=left,numberstyle=\tiny,frame=single,escapechar=',caption=Proxy as an ensemble of FSMs.\label{lst:proxyfsmens}]
'\PADU'  Proxy('\DPF'Id id, '\PADU' in'\PADU'){
    ...
    if ((frmClnt(in'\PADU')) && (isHttpRq(in'\PADU'))){'\label{lst:proxyfsmensem:req}'
       if (!cached(id, in'\PADU')){		'\label{lst:proxyfsmensem:notcache}'
             if(srvConnEstablished(id, in'\PADU'))'\label{lst:proxyfsmensem:hasconn}'
                out'\PADU'=rqstFrmSrv(id, out'\PADU');'\label{lst:proxyfsmensem:httpreq}'
             else				
                out'\PADU'=tcpSYNtoSrv(id, in'\PADU'); '\label{lst:proxyfsmensem:newconn}'
             }
        }
    /*set '\actionTags' based on context (e.g., hit/miss)*/
    out'\PADU'.'\actionTags' = ...'\label{lst:proxyfsmensem:alpha}'
    ...
    return out'\PADU';			
} 
\end{lstlisting}
\end{minipage}
\vspace{-0.6cm}
\end{figure}

\mypara{Context processing} 
The one remaining issue is the propagation and
updation of the context information in our model network. As we
saw in~\Section\ref{sec:traffic_unit}, each \DPF encodes the relevant context in the
{\tt \actionTag} field of the outgoing \PADU (Line~\ref{lst:proxyfsmensem:alpha}). 
For instance, if an \DPF modifies headers, then the \PADU encodes 
the provenance of the \PADU which can be used to check if the relevant 
policy at some downstream \DPF is implemented correctly. In summary,  
each \DPF is thus modeled as an \fsmensemble that receives an
input \PADU and generates an output \PADU with the corresponding 
updated {\tt \actionTags}.

\subsection{Network-wide modeling} 
 \label{subsec:complete_model}

Given   the per \DPF models, next we discuss how we compose these models  to
generate network-wide models.     To make this discussion concrete,  
 we use the network from Figure~\ref{fig:proxy_monitoring_policy} and see how we compose
the proxy, switch, and monitor models  in Listing~\ref{lst:network}. 

Each \DPF instance is identified by a unique id that allows us to know
the ``type'' of the \DPF and thus index into the relevant  variables.  
Lines~\ref{lst:full:switchbegin}--\ref{lst:full:switchend} model the stateless
switch. Function {\tt lookUp} takes the input \PADU, looks up its forwarding
table, and creates a new {\tt outADU} with its port value set based on the
forwarding table. Given the operators policy, parameters of the network 
and each \DPF model are configured. For example, given the policy of 
Figure~\ref{fig:proxy_monitoring_policy}, {\tt hostToWatch} is set to 
{\tt H2}. As another example, given the policy of Figure~\ref{fig:dynamic_policy},
the alarm threshold in the L-IPS is configured to 10. Following prior 
work~\cite{hsa}, we consider each switch \DPF as a static data store 
lookup updating located packets. 
Lines~\ref{lst:full:monbegin}--\ref{lst:full:monend} capture the monitoring
\DPF.  Given the actual network's topology and the library of \DPFs models, 
this composition is completely automatic and does not require the operator 
to ``write'' any code. Given the network topology \Name can  identify the 
instance of {\tt Next\_DPF} in line~\ref{lst:full:nextdfp}) of 
Listing~\ref{lst:network}.

Similar to prior work~\cite{atpg,hsa}, we consider a 
network  model in which packets are processed in a 
one-packet-per-\DPF-at-time fashion.  That is, we do not model (a)
batching or queuing effects inside the network,  (b)  parallel processing
effects inside \DPFs or  (c) simultaneous processing of different packets
across \DPFs.  Since our goal is to look for ``policy'' violations represented
in terms of \DPF-context sequences, this assumption is reasonable.  \vyas{add
to sec 3 that async/parallel effects are out of scope} 
Based on this semantics,
the data plane as a simple loop (Line~\ref{lst:full:netloop}).
 Note that because  \PADUs extend the located packet abstraction, they also
capture the locations via (e.g., {\tt networkPort} in
Listing~\ref{lst:padustruct}).  This allows \DPFs to be easily composed similar
to the composition of the simple $\Transfer$ functions~\cite{hsa}.

 In each
iteration, an \PADU is processed (Line~\ref{lst:full:packetloop}) in two steps:
(1) the \PADU is forwarded to the other end of the current link
(Line~\ref{lst:full:linkforward}) , (2) the \PADU is passed as an argument to 
the \DPF connected to this end (e.g., a switch or firewall) 
(Line~\ref{lst:full:nextdfp}) . The \PADU output by the \DPF is processed in 
the next iteration until the \PADU is ``DONE''; i.e., it either reaches its 
destination or gets dropped by a \DPF.\footnote{\seyed{pls check} Since 
\DPFs can be time-triggered (e.g., TCP connection time-out), we capture 
time using an \PADU field.  
These ``time \PADUs'' are injected by the network model periodically and upon 
receiving a time \PADU, the relevant \DPFs update their time-related state.} 
The role of {\tt assert} will become clear in the next section when we use 
symbolic execution to exercise a specific policy behavior.

\begin{figure}[t]
\vspace{-0.3cm}
\centering
\begin{minipage}[b]{210pt}
\begin{lstlisting}[frame=single,numbers=left,numberstyle=\tiny,numberblanklines=false,escapechar=',caption=Network pseudocode for Figure~\ref{fig:proxy_monitoring_policy}.\label{lst:network}]
// Symbolic '\PADUs' to be instantiated (see '\S\ref{sec:traffic-gen}').
'\PADU' A[20];
int objIdToWatch = XYZ.com
int hostToWatch = H2;
// Global state variables 
bool Cache[2][100]; // 2 proxies,  100 objects
// Model of a switch
'\PADU'  Switch('\DPF'Id id, '\PADU' in'\PADU'){ '\label{lst:full:switchbegin}'
    out'\PADU'=lookUp(id, in'\PADU');		
    return out'\PADU';			
} '\label{lst:full:switchend}'
// Model of a monitoring '\DPF'
'\PADU'  Mon('\DPF'Id id, '\PADU' in'\PADU'){'\label{lst:full:monbegin}'
     ...
     out'\PADU' = in'\PADU';
     
     if (isHttp(id, in'\PADU')){	
       takeMonAction(id, in'\PADU');/* if in'\PADU' 
       contains objIdToWatch destined to 
       hostToWatch, set out'\PADU'.dropped to 1.*/
     }
     ...		
    return out'\PADU';			
}'\label{lst:full:monend}'
// Model of a proxy '\DPF'; See Listing '\ref{lst:proxyfsmens}'
'\PADU' Proxy('\DPF'Id id, '\PADU' in'\PADU'){
   ...
}
main(){
  // Model of the data plane
  for each injected A[i]{'\label{lst:full:netloop}'
   while (!DONE(A[i])){'\label{lst:full:packetloop}'
    Forward A[i] on current link;{'\label{lst:full:linkforward}'
    A[i] = Next_DPF(A[i]);{'\label{lst:full:nextdfp}'
    '\label{lst:full:assertbegin}'assert(
    '\label{lst:full:assertprovenance}'(!(A[i].'\actionTags'[provenence]==hostId[H2]))
    '\label{lst:full:assertcached}'||(!(A[i].'\actionTags'[cacheContext]==objIdToWatch))
    '\label{lst:full:assertmonitorport}'||(!A[i].port==MonitorPort)); '\label{lst:full:assertend}'
   }
  }
}
\end{lstlisting}
\end{minipage}
\vspace{-0.6cm}
\end{figure}

\subsection{Writing future \DPF models}

 We have manually written a broad range  of \DPF models.  While we do not have
an algorithm for writing a \DPF model,  we can provide design guidelines for
writing future \DPF models based on our own methodology.  We begin by
enumerating the set of  policy scenarios  (e.g., as in the examples
of~\Section\ref{sec:motivation}); this enumeration step can be a broader
community effort in future~\cite{irtfnfv,ietf_diff_services,intel_on}. Across
the union of  these scenarios, we identify the necessary contexts (e.g., alarm,
cache hit) and corresponding \DPF states that affect these contexts (e.g., TCP
state machine of firewall, cache contents). This gives us a set of model
requirements. Then for each type of \DPF, we start with a ``dumb'' switch
abstraction and  incrementally add the logic to the model  to capture the
expected behaviors of the \DPF w.r.t.\ these required  states and contexts;
e.g., for a NAT we add per-flow consistent  mapping behaviors and  packet
provenance context.  In doing so, we make sure to identify the opportunities
for decoupling independent tasks and traffic units to enable the scalable
ensemble representation.   While we are not aware of automated tools for
synthesizing middlebox models, recent advances in program analysis and software
engineering might be a promising avenue for  automating model synthesis (e.g.,
~\cite{synoptic}).


%% file: testgenerate.tex
\section{Test Traffic Generation}
\label{sec:traffic-gen}




In this section, we describe how we use the network-wide model and 
operator's policy to generate concrete test traffic to exercise 
policy-relevant data plane states.  For \Name to be interactive for
operators, we want this step to be scalable enough to produce test plans within
seconds to a few minutes even for large networks.  Unfortunately, several
canonical search solutions ncluding model checking~\cite{clarke1999model,cbmc},
AI graph planning tools~\cite{graphplan} do not  scale beyond networks with
5-10 stateful \DPFs; e.g., model checking took  25 hours for a network with 6
switches and 3 middleboxes. Next, we describe how we make this test generation
problem tractable.

\comment
{
Recall that the  policy behavior we want to test is given to \Name as a
$(\TracePredicate;\Intention;\Action)$ specifying the subset of traffic, the
intended path through specific \DPF instances, and the eventual outcome.  
Given this policy, we  decompose the traffic generation task into two logical 
steps: (1) generating abstract test traffic (i.e., at the granularity of \PADUs) 
using symbolic execution, and (2) translating it into concrete test traffic 
(using test scripts that create test traffic). 
}


\subsection{Symbolic execution for abstract test plans}
\label{subsec:why_se}

\myparaq{Why Symbolic Execution (SE)}
 There are two key scalability concerns about test traffic generation.  
 First, we need to search  over a very large space of possible sequence 
 of traffic units.  While  \PADUs  improve scalability as compared with 
 IP packets (\Section\ref{sec:traffic_unit})  via aggregation,
we still have to search over the space of possible \PADU value  assignments.
Second, the  state space of the data plane is again very large.  While  the FSM
ensembles abstraction significantly reduces the number of states
(\Section\ref{sec:dpmodel}), it  does not address {\em state space explosion}
due to composition of \DPFs;  e.g., if the models of $\mathit{\DPF}_1$ 
and  $\mathit{\DPF}_2$ can reach $K_1$
and $K_2$ possible states for some  \PADU, respectively, the composition can
 reach $K_1 \times K_2$ states.  The traffic- and state-space explosion 
 makes  our problem (even to find abstract test traffic) challenging.

To address this scalability challenge, we turn to {\em symbolic execution} (SE),
which is a well-known approach in  formal verification to address state-space
explosion~\cite{state-exp}.  At a high level, an \SE engine
explores possible behaviors of a given program by considering different values
of \emph{symbolic variables}~\cite{acmsymbolicexecution}. 
 One  concern is that \SE  sacrifices coverage. In our specific
application context, this tradeoff to enable interactive testinr is worthwhile.
First, administrators may already have specific testing goals in mind.
Second, configuration problems affecting many users will naturally manifest
even with one test trace.  Finally, with a fast solution, we can run
 multiple tests  to improve coverage.

\mypara{Mapping policy to assertions}
For each policy scenario $(\TracePredicate;\Intention;\Action)$
 of the operator's policy (\Section\ref{sec:overview}), \Name uses \SE 
 as follows. First, we constrain the symbolic \PADUs to satisfy the 
 $\TracePredicate$ condition.  Second, we introduce the 
 negation of $\Intention$, namely
$\neg (\Intention$), as an {\em assertion} in the network model code. In
practice, given the policy and network topology, \Name instruments the network
model with $\neg(\Intention)$ assertions expressed in terms of \PADU fields
(e.g., {\tt networkPort}, {\tt \actionTags}). Then, the \SE 
engine finds an assignment to symbolic 
\PADUs such that the assertion is violated.\footnote{Note that an assertion of
the form $\neg(\mathit{A_1 \cap \dots \cap A_n})$, or equivalently 
$(\mathit{\neg A_1 \cup \dots \cup \neg A_n})$, 
is violated only if each term $A_i$ is evaluated to {\tt true}.} Because we use the
negation in the assertion, in effect, \SE concretizes a sequence of \PADUs 
that induce $\Intention$ in the network model. This abstract test traffic 
generated by \SE, after being translated into concrete test traffic and 
injected into the actual data plane, must traverse \DPFs specified in 
$\Intention$ and result in $\Action$; otherwise, the policy scenario 
is incorrectly implemented.

\mypara{Examples} To make this concrete, let us revisit 
Figure~\ref{fig:proxy_monitoring_policy} in Listing~\ref{lst:network}, 
where  we want a test plan to observe cached responses from 
the proxy to Dept. Lines~\ref{lst:full:assertbegin}-\ref{lst:full:assertend} 
show the assertion  to get a trace (i.e., a sequence of \PADUs) that 
change the state  of the data plane such that the last \PADU in the abstract
traffic trace: (1) is from host $H_2$ (Line~\ref{lst:full:assertprovenance}),
(2) corresponds to a cached respnose (Line~\ref{lst:full:assertcached}),
 and (3) reaches the network port where the monitor is attached  to 
 (Line~\ref{lst:full:assertmonitorport}). For example, the  \SE  engine 
 might give us a test plan with 5 \PADUs: three \PADUs between a 
 host in the Dept. and the proxy to establish a TCP connection (the 
 3-way handshake), a fourth \PADU has {\tt httpGetObj = httpObjId} 
 from the host to the proxy (a cache miss), followed by another \PADU 
 with the field {\tt httpGetObj} set to {\tt httpObjId} to induce a cached 
 response.  Similarly, Listing~\ref{lst:ips_assert} shows an assertion 
 in  Lines~\ref{lst:ups:ips_assert_begin}-\ref{lst:ups:ips_assert_end}
 so that an alarm is triggered at both L-IPS and H-IPS of the example 
 from Figure~\ref{fig:dynamic_policy}.

\begin{figure}[t]
\vspace{-0.3cm}
\centering
\begin{minipage}[b]{200pt}
\begin{lstlisting}[frame=single,numbers=left,numberstyle=\tiny,numberblanklines=false,escapechar=',caption=Assertion pseudocode for Figure~\ref{fig:dynamic_policy} to trigger alarms at both IPSes.\label{lst:ips_assert}]
// Global state variables
int L_IPS_Alarm[noOfHosts];//alarm per host
int H_IPS_Alarm[noOfHosts];//alarm per host
...
    '\label{lst:ups:ips_assert_begin}'assert((!L_IPS_Alarm[A[i].srcIP]) || 
    	   (!H_IPS_Alarm[A[i].srcIP]));	'\label{lst:ups:ips_assert_end}'
\end{lstlisting}
\end{minipage}
\vspace{-0.6cm}
\end{figure}

\subsection{Optimizing \SE}
\label{subsec:se_opt}

While \SE is orders of magnitude faster than other candidates
as the search mechanism, it is still not sufficient for interactive 
testing;  even after a broad sweep of configuration parameters 
and command line arguments (e.g., {\tt max-sym-array-size}, 
{\tt max-memory}, and {\tt optimize}) 
to customize \klee, it took several hours even for a small topology 
(\Section\ref{sec:eval}).  To  scale to larger topologies, 
we implement two key optimizations:

\begin{packeditemize}

\item {\em Minimizing number of symbolic variables:} Making an entire 
\PADU  structure (Listing~\ref{lst:padustruct}) symbolic will force 
\klee to find values for every field. To avoid   this, \Name uses the
policy scenario to determine a small subset of \PADU fields as 
symbolic; e.g., when it is testing data plane with a stateful firewall 
but without a proxy, it makes the HTTP-relevant fields concrete 
(i.e., non-symbolic) by assigning the \emph{don't care value} 
$*$ (represented by -1 in our implementation) to them. As another 
example, \Name  sets a client's TCP port number to a temporary 
value (as opposed to   making the {\tt srcPort} field symbolic). This 
value is only used in   the model for test planning and the actual 
client TCP port is chosen  by the host at run 
time (\S\ref{subsec:concrete})

 

\item {\em Scoping values of symbolic variables:} The
  $\TracePredicate$ already scopes the range of values each \PADU can
  take. \Name further narrows this range by using the policy scenario to 
  constrain possible  values of symbolic \PADU fields. For example, while
 {\tt tcpSYN} is an integer \PADU field, \Name restricts its value to be
 either 0 or 1 to shrink the search space.
 
 
%

\end{packeditemize}

\subsection{Generating concrete test traffic}
\label{subsec:concrete}

The output of \SE is a sequence of \PADUs $\mathit{\PADUSeq}^\mathit{\SE}$, 
and our next goal is to translate it into concrete test packets. Since \PADUs 
are abstract I/O units, we cannot  directly inject them into the data plane. Moreover,  
we cannot simply do a  one-to-one translation between \PADUs and  raw 
packets and do a trace replay~\cite{bit-twist}; e.g., we need some session 
semantics for TCP or in an actual HTTP session several
parameters will be outside of our control (e.g., chosen by the remote server at
the test run time).  While we do not claim to have a comprehensive algorithm  
for translating an arbitrary $\mathit{\PADUSeq}^\mathit{\SE}$ into concrete 
test traffic, we use a heuristic approach as follows.  

We have created a {\em library} using domain knowledge  to map
a known  $\mathit{\PADUSeq}_\testscriptindex$ into {a \em test script}.
For instance, if we have an \PADUSeq
consisting of three \PADUs for TCP connection establishment and a web request,
we map this into a simple {\tt wget} request with the required parameters
(e.g., server IP and object URL) for the request indicated by the \PADUSeq.
 In the most basic  case, the script will be a simple IP  packet.
In our current implementation,  we have manually populated this library and
currently use 11 such traffic generation primitive functions (e.g.,
\texttt{closeTCP(.)}, \texttt{getHTTP(.)}, \texttt{sendIPPacket(.)}) that
support IP, TCP, UDP, HTTP, and FTP. Automating the task of populating
such a trace library is outside the scope of the paper.

 Now, given a $\mathit{\PADUSeq}^\mathit{\SE}$, we use this library as follows.
We partition the $\mathit{\PADUSeq}^\mathit{\SE}$ based on srcIP-dstIP
pairs (i.e., communication end-points) of \PADUs; i.e.,  $\mathit{\PADUSeq}^\mathit{\SE} =
\bigcup_\testscriptindex  \mathit{\PADUSeq}_\testscriptindex$.  Then for each
partition  $\mathit{\PADUSeq}_\testscriptindex$, we do a
longest-specific match (i.e., match on a  protocol at the highest possible layer 
of the network stack) in our test script library, retrieve the corresponding
{\testscript}s for each subsequence  and then  concatenate  these scripts.
We acknowledge this step is heuristic and creating a comprehensive 
 mapping process is outside the scope of this paper.

%% file: validation.tex
\section{Test Monitoring and Validation}
\label{sec:validation}

After the test traffic is injected into the data plane, the outcome
should be monitored and validated.  First, we need to disambiguate
true policy violations from those caused by background interference.
Second, we need mechanisms to help localize the misbehaving \DPFs.
While a full solution to fault diagnosis and localization is outside
the scope of this paper, we discuss the practical heuristics we
implement.

\mypara{Monitoring} Intuitively, if we can monitor the status of the
network in conjunction with the test injection, we can check if any of
the background or non-test traffic can potentially induce false policy
violations.  Rather than monitor all traffic (we refer to this as
\allPortsMon), we can use the intended policy to capture a smaller
relevant traffic trace; e.g., if the policy is involves only traffic
to/from the proxy, then we can focus on the traffic on the proxy's
port.  To further minimize this monitoring overhead, as an initial
step we capture relevant traffic only at the switch ports that are
connected to the stateful \DPFs rather than collect traffic traces
from all network ports. However, if this provides limited visibility
and we need a follow-up trial (see below), then we revert to logging
traffic at all ports for the follow-up exercise.
 
\mypara{Validation and localization} Next, we describe our current
workflow to validate if the test meets our policy intent, and (if the
test fails) to help us localize the sources of failure otherwise.  The
workflow naturally depends on whether the test was a success/failure
and whether we observed interfering traffic as shown in
Table~\ref{tab:valProc}.

 \begin{table}[t]
  \begin{center}
  \begin{footnotesize}
  \begin{tabular}{p{2.8cm}|p{2cm}|p{2cm}}
             	  		& $\mathit{Orig}=\mathit{Obs}$  		&  $\mathit{Orig} \neq \mathit{Obs}$	\\ \hline
         No interference or resolvable interference       	&  Success  &  Fail. Repeat on  $\mathit{Orig}-\mathit{Obs}$ using \allPortsMon   \\    \hline 
         Unresolvable interference &   \multicolumn{2}{c}{Unknown;  Repeat  $\mathit{Orig}$ using \allPortsMon}
     \end{tabular}
  \end{footnotesize}
  \end{center}
  \vspace{-0.4cm}
 \tightcaption{Validation and test refinement workflow.}
  \vspace{-0.4cm}
 \label{tab:valProc}
 \end{table}

Given the specific \intent we are testing and the relevant traffic
logs, we determine if the network satisfies the intended behavior;
e.g., do packets follow the policy-mandated paths?  In the easiest
case, if the observed path $\mathit{Obs}$ matches our intended
behavior $\mathit{Orig}$ and we have no interfering traffic, this step
is trivial and we declare a success.  Similarly, if the two paths
match, even if we have potentially interfering traffic, but our
monitoring reveals that it does not directly impact the test (e.g., it
was targeting other applications or servers), we declare a success.

Clearly, the more interesting case is when we have a test failure;
i.e., $\mathit{Obs} \neq \mathit{Orig}$.  If we identify that there
was no truly interfering traffic, then there was some potential source
of policy violation.  Then we identify the largest common path prefix
between $\mathit{Obs}$ and $\mathit{Orig}$; i.e., the point until
which the observed and intended behavior match and to localize the
source of failure, we zoom in on the ``logical diff'' between the
paths. However, we might have some logical gaps because of our choice
to only monitor the stateful \DPF-connected ports; e.g., if the proxy
response is not observed by the monitoring device, this can be because
of a problem on any link or switch between the proxy and the
monitoring device.  Thus, when we run these follow up tests, we enable
\allPortsMon to obtain full visibility.

Finally, for the cases where there was indeed some truly interfering
traffic, then we cannot have any confidence if the test
failed/succeeded even if $\mathit{Obs} = \mathit{Orig}$. Thus, in this
case the only course of action is a fall back procedure to repeat the
test but with \allPortsMon enabled.  In this case, we use an
exponential backoff to wait for the interfering flows to die.

%% file: implementation.tex
\section{Implementation}
\label{sec:implementation}
\seyed{TODO: say something about modeling time with a special \PADU 
field? (since we are talking about timeouts in the example fsms)}


\myparatight{\DPF models} We wrote C models for switches, ACL devices,
stateful firewalls (capable of monitoring TCP connections and blocking
based on L3/4 semantics), NATs, L4 load balancers, HTTP/FTP proxies, 
passive monitoring, and simple intrusion prevention systems ( 
counting failed connection attempts and matching payload signatures).
 Our models are between 10 (for a switch) to 100 lines (for 
a proxy cache) of C code. The main loop of network model, utility 
functions, and header files (e.g., \PADU definitions and utility functions) 
have a total of fewer than 200 LoC. To put these numbers in context, the 
 real-world middleboxes can range from 
 2K (e.g., Balance~\cite{balance}) to few 100K (e.g., Squid~\cite{squid}, 
Snort~\cite{snort}).   We reuse common templates 
  across \DPFs;  e.g., TCP connection sequence used both in 
firewall model and proxy model.

\myparatight{Validating \DPF models}  First, we use a bounded model checker,
\cbmc~\cite{cbmc}, on individual \DPF models and the network model to ensure
they do not contain software bugs (e.g., pointer violations). This was
a time-consuming but one-time task. Second, we used call graphs
visualization~\cite{valgrind, kcachegrind} based on extensive, manually generated
input traffic traces to check that the model behaves as expected.

\myparatight{Test traffic generation and injection} We use \klee with
the optimizations discussed earlier to produce the \PADU-level test 
traffic, and then translate it to test scripts that are deployed at the
injection points. Test traffic packets are marked by setting a specific 
(otherwise unused) bit.

\myparatight{Traffic monitoring and validation} 
We currently use offline monitoring via {\tt tcpdump} (with 
suitable filters); we plan to integrate more real-time solutions 
like NetSight~\cite{netsight}.  We use  
{\tt OpenFlow}~\cite{openflow} to poll/configure switch state.

%% file: evaluation_new.tex
\section{Evaluation}
\label{sec:eval}

In this section,  we show that: \\
 (1) \Name  enables close-to-interactive running 
times even for large topologies (\Section\ref{sec:eval_performance}); \\ 
 (2) \Name's design is critical for scalability 
(\Section\ref{sec:design_choices_effect}); and \\
(3) \Name successfully helps diagnose 
a broad spectrum of data plane policy violations 
(\Section\ref{sec:eval_effectiveness}).

\myparatight{Testbed and topologies} To run realistic large-scale experiments
with large topologies, we use a testbed of 13 server-grade machines (20-core
2.8GHz servers with 128GB RAM) connected via a combination of direct 1GbE links
and a 10GbE Pica8 OpenFlow-enabled switch.  On each server, with KVM installed,
we run injectors and  software \DPFs  as separate VMs, connected via {\tt
OpenvSwitch} software switches. The specific stateful \DPFs (i.e., middleboxes)
are iptables~\cite{iptables} as a NAT and a stateful firewall, Squid~\cite{squid} 
as a proxy, Snort~\cite{snort} as an IPS/IDS, Balance~\cite{balance} as the load 
balancer, and PRADS~\cite{prads} as a passive monitor.

In addition to the example scenarios from
\Section\ref{sec:motivation},  we use 8 randomly selected recent  topologies
from the Internet Topology Zoo~\cite{topology_zoo} with 6--196 nodes. We also
use two larger topologies (400 and 600 nodes)  by extending these topologies.
These serve as  switch-level  topologies; we  extend them with different \DPFs
to enforce policies. As a concrete policy  enforcement scheme  we
implemented a tag-based solution to handle dynamic 
middleboxes~\cite{flowtags_nsdilong}. 
 We reiterate that the design/implementation of this scheme is not the goal 
 of \Name; we simply needed {\em some} concrete solution.


\subsection{Scalability of \Name}
\label{sec:eval_performance}

We envision operators using \Name in an interactive fashion; i.e., the time for
test generation should be within 1-2 minutes even for large networks with
hundreds of switches and middleboxes. 


\mypara{Impact of topology size} We fix the 
policy size (i.e., the length the chain of stateful \DPFs in the policy) to 3, 
including a NAT, followed by a proxy, followed by a stateful firewall. 
The firewall is expected to block access from a fixed subset
of origin hosts to certain web content. To each switch-level topology, 
 we add  a  number of middleboxes (0.5$\times$ \#switches)  and connect each middlebox to a randomly
 selected switch with  at most one middlebox connected to each
switch. There is also one host connected to each switch that will be used
as the end point of policies. The smallest topology with 6 switches 
(Heanet) has one instance of the policy chain (i.e.,  a NAT, a proxy, and 
a firewall);  we linearly increase the number of policy chains 
 to test as a function of topology size.

\begin{figure}[t]
  \centering
  \includegraphics[width=220pt]{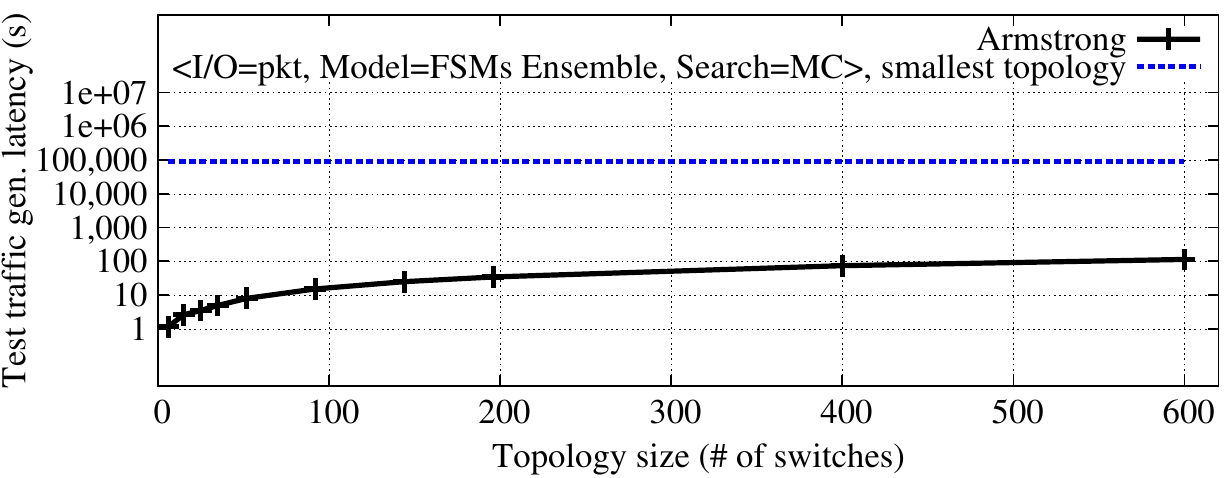}
  \tightcaption{Test generation latency vs.\ topology size.}
  \label{fig:new_topology_size_effect}
\end{figure}

Figure~\ref{fig:new_topology_size_effect} shows the average test traffic
generation latency. (Values are close to the average we do not show error
bars). In the largest topology with 600 switches and 300 middleboxes (i.e., 100
policy chain instances), the traffic generation latency of \Name is 113
seconds. To put this  in context, we also show the traffic generation time of a
strawman solution of using \cbmc~\cite{cbmc} model checker on our data plane
model.  Even on a tiny 9 node topology with 6 switches and 3 middleboxes this
took 25 hours; i.e., \Name on 90$\times$ larger 
topology is {\em at least} five orders of faster than the status quo.
 Note that this result considers \Name running sequentially;
 we can trivially 
 parallelize \Name across the different policy scenarios. 
 

\begin{figure}[th]
  \centering
  \includegraphics[width=220pt]{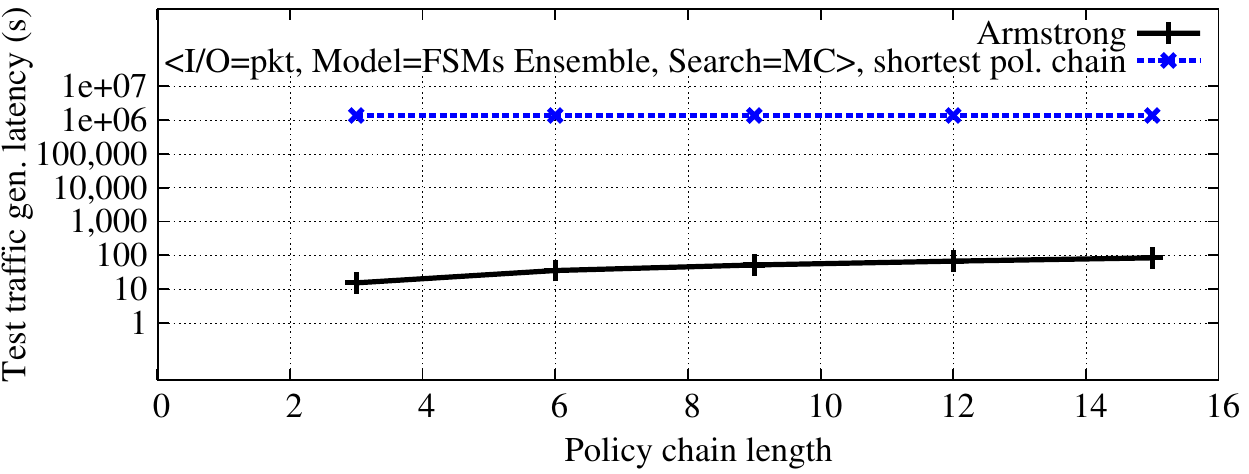}
  \tightcaption{Test latency vs.\ policy chain length.}
  \label{fig:new_policy_size_effect}
\end{figure}

\mypara{Impact of policy complexity}  Next we consider the effect of policy
complexity measured by the number of middleboxes present in the policy. 
We fix the topology to have 92 switches (OTE GLOBE). To stress
test \Name, we generate synthetic longer chains in which the intended action of
each \DPF on the chain depends on some contextual information from the previous
\DPF.  Figure~\ref{fig:new_policy_size_effect} shows that even in case of the
longest policy chain with 15 middleboxes, \Name  takes only 84 seconds. 
 Again to put the number in context we show the strawman.


\mypara{Break-down of test traffic generation latency} Recall 
from~\Section\ref{sec:traffic-gen} that test generation in \Name 
has two stages. We find that translating abstract test traffic
into concrete test traffic composes between 4--6\% of the entire 
latency to generate the test traffic; this is the case across different
topology sizes and policy sizes (i.e., policy chain lengths) (not shown).

\mypara{End-to-end overhead} After \Name generates 
test traffic, it injects the test traffic, monitors it, and determines 
the result. The actual test on the wire lasts $\leq$  3 seconds 
in our experiments with 600 switches and 300 middleboxes. 
 However, we monitor the network for a longer 10-second window 
 to capture possibly relevant traffic  events.  On our largest topology 
 with 600 nodes this validation analysis took only 87 seconds (not shown).




\comment
{
 \begin{table}[t]
  \begin{center}
  \begin{footnotesize}
  \begin{tabular}{p{1.4cm}|p{.7cm}|p{1.2cm}|p{0.8cm}|p{0.7cm}|p{0.7cm}}
         Topo.(\# of switches)     &  Cwix (35)  & OTEGlobe (92) 	&Cogent (196) & 400-node & 600-node\\ \hline
          Time (s)               &   3.8  & 	 9.9     &   25.2 & 56.3 & 87.6   \\
     \end{tabular}
  \end{footnotesize}
  \end{center}
  \vspace{-0.4cm}
 \tightcaption{Validation latency for different topologies.}
 \label{tab:validation}
 \end{table}
}
\subsection{ \Name design choices}
\label{sec:design_choices_effect}

Next, we do a component wise analysis to demonstrate the effect of our key
design choices and optimizations.

\mypara{Code vs.\ models}  Running \klee on  smallest 
 \DPF codebase of around 2000 LOC (i.e., balance~\cite{balance}) took about 20 
hours. In very small experiment with policy chain of length 2 involving 
only one switch directly connected to a client, a server, a load balancer, 
and a monitor~\cite{prads}, traffic generation time took 57 
hours (not shown).

\comment
{
 \begin{table}[t]
  \begin{center}
  \begin{footnotesize}

  \begin{tabular}{p{2.6cm}|p{2.6cm}|p{2.3cm}}
         			 &  Load balancer (LB)	& LB+Monitor\\ \hline
          \klee on ``raw'' code  &   20 (hrs)  & 	 57 (hrs)        \\ \hline
          \Name  &   $<$ 1 sec  & 	 $<$ 1 sec      \\
     \end{tabular}
  \end{footnotesize}
  \end{center}
  \vspace{-0.4cm}
 \tightcaption{\klee on \Name model  vs.\ \DPF code}
 \label{tab:se_on_code}
 \end{table}
}

\begin{figure}[th]
  \centering
  \includegraphics[width=230pt]{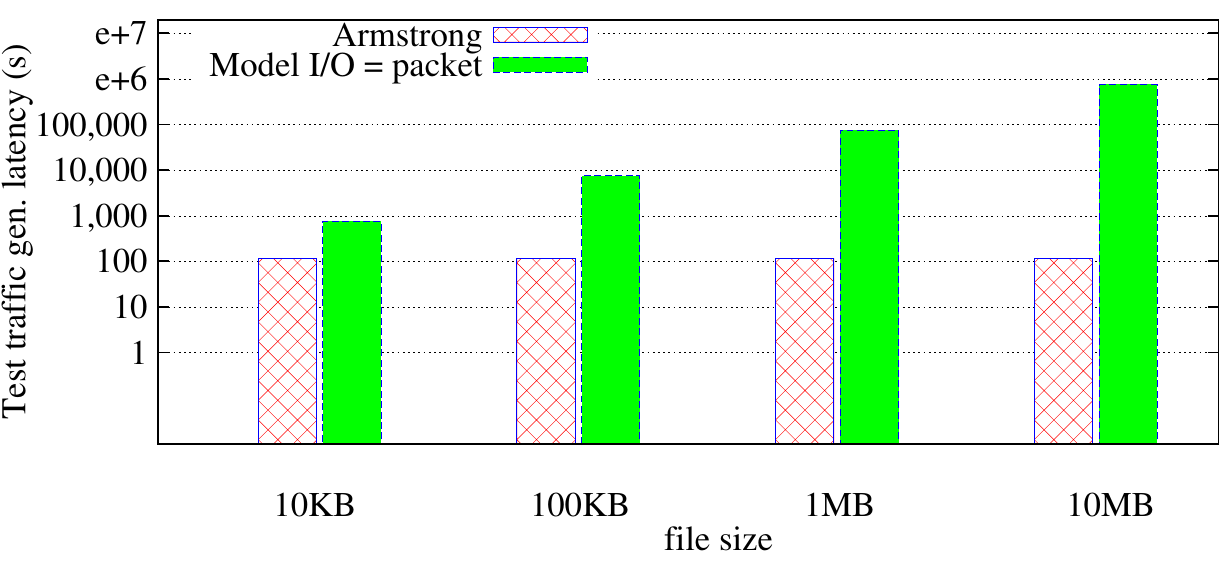}
  \tightcaption{Effect of using \PADU vs.\ packet  for various request sizes.} 
  \label{fig:adu_effect}
\end{figure}

\mypara{\PADU vs.\ packet} First, to see how aggregating 
a sequence of packets as an \PADU helps with scalability, we 
 vary file size in  an HTTP request and response scenario. 
 Then,  we use \Name to generate test traffic to test the proxy-monitor 
 policy (Figure~\ref{fig:proxy_monitoring_policy}) in terms of \PADUs vs.\ raw 
 MTU-sized packets. Figure~\ref{fig:adu_effect} shows  that 
 on  the topology with 600 switches and 300 middleboxes test traffic generation latency
 increases \vyas{linearly}  vs.\ the size of the response. Because  the 
 number of test packets is dominated by the number of object  
retrieval packets, aggregating all file retrieval packets as one \PADU 
significantly cuts the latency of the test traffic generation.
(The results, not shown, are  consistent across topologies as well
as using FTP instead of HTTP.)


%

\mypara{SE vs.\ model checking} Our results already showed
dramatic gains of \Name w.r.t.\ model checking on the raw code.  One natural
question is if model checking could have benefited from the other \Name
optimizations.  To this end, we evaluated the performance of an optimized
\cbmc-based model checking solution with \Name-specific optimizations such as
FSM Ensembles, \PADUs, and other scoping and variable reduction optimizations
(\Section\ref{sec:traffic-gen}).  This optimized version was indeed
significantly faster than before but it was still  two orders of magnitude
slower than \Name (not shown).  This suggests that while  our abstractions are
independently  useful for other network verification efforts using model
checking, these mechanisms are not directly suitable for the interactive 
testing time-scales  we envision in \Name.

\begin{figure}[t]
  \centering
  \includegraphics[width=230pt]{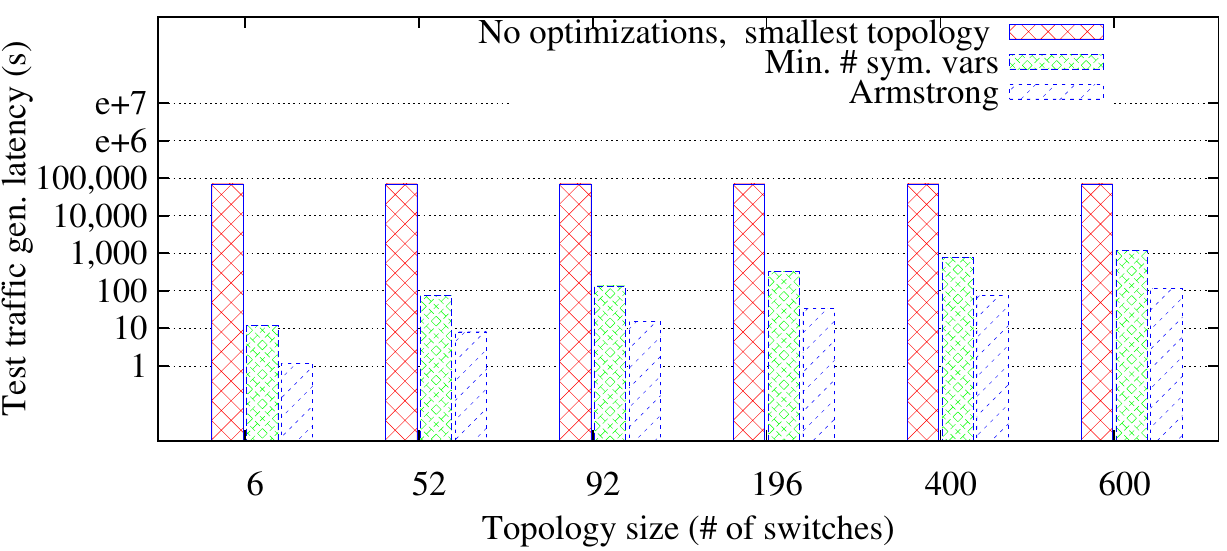}
  \tightcaption{Improvements due to  \SE optimizations.}
  \label{fig:opt_effect}
\end{figure}

\mypara{Impact of  SE optimizations} We examine the effect of the
SE-specific optimizations (\Section\ref{sec:traffic-gen})  in
Figure~\ref{fig:opt_effect}. To put our numbers in context, using \klee
without any  optimizations on a network of six switches and one
policy chain  with three middleboxes took $\geq$ 19 hours.
 We see that  minimizing  the number of symbolic variables reduces 
 the test generation latency   by  three orders of magnitude and  
 scoping the values  yields a further $>9\times$  reduction. 

\vyas{dont do linear projection}

\subsection{End-to-end use cases}
\label{sec:eval_effectiveness}

Next we also demonstrate the effectiveness of \Name in finding 
policy violations.

\mypara{Diagnosing induced enforcement bugs}  We used  a ``red team--blue team'' 
 to evaluate the end-to-end utility of \Name in debugging policy violations. 
  Here, the red team (Student 1) informs of the blue team
(Student 2) of \intents for each network, and then secretly picks one of 
the intended behaviors (at random) and creates a failure mode that 
causes the network to violate this \intent; e.g., misconfiguring the 
L-IPS count threshold or disabling some control module.  The blue 
team uses \Name to (a) identify that a violation occurred and (b) 
localize the source of the policy violation.  We also repeated these 
experiments reversing  student roles; but do not show these 
results for brevity.

\begin{table}[t]  
\begin{small}
\begin{center}
    \begin{tabular}{ p{4cm} | p{4cm} }
    {\bf ``Red Team'' scenario} &  {\bf \Name test trace} 
    \\ \hline
Proxy/Mon (Fig.~\ref{fig:proxy_monitoring_policy}); $S_1$-$S_2$ link is down &  Non-cached rqst from inside the Dept, followed by request for the same object from by another source host in the Dept\\ \hline
Proxy/Mon (Fig.~\ref{fig:proxy_monitoring_policy}); The port of $S_1$ connected to proxy is misconfigured to not support OpenFlow & HTTP rqst from Dept	\\ \hline
 Cascaded NATs  (Fig.~\ref{fig:cascaded_nats}); {\tt FlowTags}~\cite{flowtags_nsdilong} controller  shut down &  $H_1$ attempts to access to the server \\ \hline
 Multi-stage triggers (Fig.~\ref{fig:dynamic_policy}); L-IPS miscounts by summing three hosts  & $H_1$ makes 9  scan attempts followed by 9 scans by $H_2$  \\ \hline
Conn. limit.; Login counter resets  &  $H_1$ makes 3 continuous log in attempts with a wrong password   \\ \hline
Conn. limit.; $S_1$ missing switch forwarding rules from AuthServer to the protected server  & $H_2$ makes a log in attempt with the correct password
\end{tabular}
\end{center}
\end{small}
\vspace{-0.2cm}
\tightcaption{Some example red-blue team scenarios.}
\label{tab:bug_scenarios}
\vspace{-0.2cm}
\end{table}

Table~\ref{tab:bug_scenarios} highlights the results for a subset of
these scenarios and also shows the specific traces that \Name 
generated.  Three of the scenarios use the motivating examples
from \Section\ref{sec:motivation}. In the last scenario
(Conn. limit.), two hosts are connected to a server through an
authentication server to prevent brute-force password guessing
attacks. The authentication server is expected to halt a host's access
after 3 consecutive failed log in attempts.  In all scenarios the
blue-team successfully localized the failure (i.e., which \DPF,
switch, or link is the root cause) within 10 seconds. Note that these
bugs could not be exposed with existing debugging tools such as
ATPG~\cite{atpg}, ping, or traceroute.\footnote{They can detect
   link/switch failure being down but cannot
  capture subtle bugs w.r.t.\ stateful/context-dependent behaviors.}

\myparatight{Loops and reachability} \Name can also  help in diagnosis
reachability problems as well.  It is worth nothing that while checking such
properties in stateless is easy~\cite{hsa}, this does not extend to stateful
data planes. We extended \Name  to support reachability properties via new
use of assertions. For instance, to detect loops we add  assertions of the form: {\tt
assert(seen[ADU.id][port]<K)}, where {\tt ADU} is a symbolic \PADU, {\tt port} is
a switch port, and {\tt K} reflects a simplified definition of a loop that  the
same \PADU is observed at the same port $\geq$ {\tt K} times. Similarly, to
check if some traffic  can reach  {\tt PortB} from {\tt PortA} in the network,
we initialize a \PADU with the port field  to be {\tt PortA} and use an
assertion of the form {\tt assert(\PADU.port != PortB)}.  Using this technique
we were able to detect synthetically induced switch forwarding loops in
stateful data planes (not shown).

%% file: relwork.tex
\section{Related Work}
\label{sec:related}


\myparatight{Network verification} There is a rich literature on static
reachability checking~\cite{margrave,fireman,maltzinfocom,hsa,netplumber,anteater}.
At a high level, these focus on simple properties (e.g., black holes, loops) and do not tackle 
 networks with complex middleboxes.  NICE combines model checking  and symbolic execution to find bugs
in  control plane software~\cite{nice}. \Name is complementary in that it
generates test cases for data plane behaviors.  Similarly, SOFT generates
  tests to check  switch implementations against a specification~\cite{soft}.
Again, this cannot be extended to middleboxes.

\myparatight{Test packet generation}  The work closest in spirit to \Name is
ATPG~\cite{atpg}, which builds on  HSA to generate test packets to test
reachability. As we discussed earlier~\ref{sec:motivation}, it cannot be applied to
our scenarios.  First, middlebox behaviors  are not ``stateless transfer
functions'', which is critical for the scalability of ATPG. Second, the
behaviors we want to test require us to look beyond single-packet test cases. 

\myparatight{Programming languages} Other work attempts to generate
``correct-by-construction'' programs~\cite{frenetic,vericon,netkat}.  Currently 
  their semantics  do not currently capture  {\em stateful}
data planes and {\em context-dependent} behaviors. 
 That said,   our work in \Name is complementary to such enforcement mechanisms; 
 e.g., active testing may be our only option to check if the network with proprietary \DPFs behaves as intended.

\myparatight{Network debugging}  There is a rich literature for fault
localization in networks and systems
(e.g.,~\cite{gestalt,pip,xtrace,mcs}).  These algorithms can  be
used in  the inference engine of \Name. Since this is not the primary focus
of our work, we use simpler heuristics.

\myparatight{Modeling middleboxes} Joseph and Stoica formalized middlebox
forwarding behaviors but  don't model stateful behaviors~\cite{joseph_mbox}.
The only work that models stateful behaviors are
FlowTest~\cite{flowtest_hotsdn}, Symnet~\cite{symnet}, and  work by Panda
et~al~\cite{pandaarxiv}.  FlowTest's high-level models are not composable and
the AI planning approaches do not scale beyond 4-5 node networks.
Symnet~\cite{symnet} uses models written in Haskell to capture NAT semantics
similar to our example; based on published work we do not have details on their
models, verification procedures, or scalability.  The work of Panda et~al.,  is 
 different from \Name both in terms of  goals (reachability and isolation) and  
techniques (model checking).


\myparatight{Simulation and shadow configurations} Simulation~\cite{ns3},
emulation~\cite{mininet,emulab},   and shadow configurations~\cite{shadow} are
common methods to model/test networks.  \Name is orthogonal in 
that it focuses  on {\em generating test scenarios}. While our current focus  is 
on active  testing,  \Name's  applies to  these platforms as well.
 We also posit that our 
 techniques  can be used to validate these efforts.

\seyed{removed from model section to downplay flowtags there: 
 There is a natural correspondence 
between {\tt \actionTags} and {\tt
  FlowTags} \vyas{downplay flowtags and move to relwork} 
we used previously to track packet modifications and
dynamic middlebox actions~\cite{flowtags_nsdilong}.

Note that \Name does not require that the actual \DPFs be
{\tt FlowTags}-enabled; it merely uses these {\tt FlowTags}-like
constructs internally to model \DPF operations.
}

%% file: conclusions.tex
\section{Conclusions}
\seyed{say something like if we are to go toward CAD for networks, it is
natural to bring in insights from software/hardware engineering that have been
doing this for decades \dots}

\Name tackles a key missing piece of existing network verification
efforts---context-dependent policies and
stateful data planes introduce fundamental expressiveness and scalability
challenges for existing abstractions and exploration mechanisms.  We make 
 three key contributions to address these challenges: (1) a novel
\PADU abstraction for modeling network I/O behavior; (2) tractable modeling of
\DPFs as  \fsmensembles; and (3) an optimized test workflow using
symbolic execution.  We demonstrate that \Name can handle complex policies over
large networks with hundreds of middleboxes within 1-2 minutes.  In doing so
we take the ``CAD for networks'' vision one  step closer to reality. 
